\documentclass[fleqn,usenatbib]{mnras}
\usepackage{lipsum}

\usepackage{newtxtext,newtxmath}

\usepackage[T1]{fontenc}
\usepackage{lipsum}

\DeclareRobustCommand{\VAN}[3]{#2}
\let\VANthebibliography\thebibliography
\def\thebibliography{\DeclareRobustCommand{\VAN}[3]{##3}\VANthebibliography}

\usepackage{graphicx}	
\usepackage{amsmath}	
\usepackage{subfig}
\usepackage{float}

\title[Doppler Tomography of HD 179949 b]{Doppler Tomography as a tool for characterising exoplanet atmospheres II: an analysis of HD 179949 b}

\author[S. M. Matthews et al.]{S. M. Matthews,$^{1}$
C. A. Watson,$^{1}$\thanks{E-mail: c.a.watson@qub.ac.uk (CAW)} E. J. W. de Mooij,$^{1}$, T. R. Marsh$^{2}$,  M. Brogi$^{2,3}$,\newauthor S. R. Merritt$^{1}$, K. W. Smith$^{1}$, D. Steeghs$^{2}$
\\
$^{1}$Astrophysics Research Centre, Queen’s University Belfast, Belfast BT7 1NN, UK\\
$^{2}$Department of Physics, University of Warwick, Coventry CV4 7AL, UK\\
$^{3}$Department of Physics, University of Turin, Via Pietro Giuria 1, I-10125, Turin, Italy\\}

\date{Accepted XXX. Received YYY; in original form ZZZ}

\pubyear{2022}

\begin{document}
\label{firstpage}
\pagerange{\pageref{firstpage}--\pageref{lastpage}}
\maketitle

\begin{abstract}
High-resolution Doppler spectroscopy provides an avenue to study the atmosphere of both transiting and non-transiting planets. This powerful method has also yielded some of the most robust atmospheric detections to date. Currently, high-resolution Doppler spectroscopy  detects atmospheric signals by cross-correlating observed data with a model atmospheric spectrum. This technique has been successful in detecting various molecules such as H$_{2}$O, CO, HCN and TiO, as well as several atomic species. Here we present an alternative method of performing high-resolution Doppler spectroscopy, using a technique known as Doppler tomography. We present an analysis of HD 179949 b using Doppler tomography and provide Doppler tomograms confirming previous detections of CO at 2.3 \micron\, and H$_2$O at both 2.3 \micron\, and 3.5 \micron ~within the atmosphere of HD 179949 b, showing significantly lower background noise levels when compared to cross-correlation methods applied to the same data. We also present a novel detection of H$_{2}$O at 2.1 \micron\, as well as a tentative detection of CO on the night side of the planet at 2.3 \micron. This represents the first observational evidence for molecular absorption in the night-side emission spectrum of an exoplanet using Doppler spectroscopy.
\end{abstract}




\section{Introduction}
\label{sec:intro}

The study of exoplanet atmospheres allows us to determine their chemical composition and fundamental conditions (such as temperature and pressure) that, in turn, may also shed light on the planetary formation and subsequent evolution. In addition, this allows us to probe the physics driving key observed characteristics of exoplanet atmospheres themselves and thereby constrain underlying theories posed to explain these properties \citep[e.g.][etc.]{Fortney2008, Madhusudhan2012, Burrows2001}.

One of the key methods employed to observationally study exoplanet atmospheres is high-resolution Doppler spectroscopy. This technique takes advantage of the fact that the spectral features from the planet are Doppler-shifted due to its orbital motion. Since hot Jupiters orbit their host star at high velocities (typically hundreds of km s$^{-1}$), while the stellar reflex Doppler motion is orders of magnitude smaller, the stellar and planetary features are significantly separated in velocity over the majority of the planet's orbit. Furthermore, since telluric lines are also stationary with respect to the observer, they can also be removed without significantly affecting any underlying Doppler shifting planetary signal. By measuring the Doppler shift of both the planet and the star it is also possible to determine model-free masses for both components as the system can be effectively treated as a double-lined spectroscopic binary. This is particularly useful for non-transiting planetary systems, since it enables the true planetary and stellar masses to be determined for these systems.

In general, however, the planetary signal is too weak to be detected using individual lines. Therefore, the signal has to be extracted by combining a multitude of lines from the planetary atmosphere. Typically, this is done by cross-correlating the data with a model atmosphere. Since these atmospheric models target specific molecular and/or atomic species, there is little ambiguity in the nature of detections relative to low-resolution observations, as systematics are extremely unlikely to match the exact high-resolution spectral fingerprint of a species, as opposed to the broadband features seen at low-resolution.

High resolution Doppler spectroscopy was first successfully implemented by \cite{Snellen2010} to robustly detect CO in the transmission spectrum of the hot-Jupiter HD 209458 b. Here they also found evidence for winds blowing from the day- to the night-side in the form of a blue-shift of the CO signal relative to the planetary system's systemic velocity, highlighting the capability of such techniques to also uncover planetary atmosphere dynamics. Subsequent high-resolution studies have also detected the thermal emission from near-infrared observations of the day-side of hot-Jupiters (e.g. \citealt{Brogi2012}, \citealt{deKok2013}). In addition, many molecular species beyond CO have been detected, including $\mathrm{H_{2}O}$ \citep[e.g.][]{Birkby2013, Birkby2017}, HCN \citep[e.g][]{Cabot2019}, CH$_4$ \citep[e.g.][]{Guilluy2019, Giacobbe2021}, $\mathrm{NH_{3}}$ \citep[e.g.][]{Giacobbe2021, Carleo2022, Guilluy2022} and $\mathrm{C_{2}H_{2}}$ \citep[e.g.][]{Giacobbe2021, Guilluy2022}. Observations in the optical have also yielded evidence of many neutral atomic species that include, but are not limited to, Fe I, Mg I, Na I, Ca I, Cr I, Mn I, Ti I, V I and Ni I \citep[e.g.][]{Hoeijmakers2019, Hoeijmakers2020, Gibson2020, Cabot2020, Yan2020, Turner2020, Merritt2021, Herman2022, Cont2022, Kesseli2022} as well as several ionised species \citep[e.g.][]{Merritt2021, Cont2022, Silva2022}. While evidence for other molecules, such as TiO \citep[e.g.][]{Cont2021, Nugroho2017}, has been reported at high resolution, many other attempts to find these species \citep[e.g.][]{Hoeijmakers2015, Merritt2020} have proved unsuccessful. This could be due to either an inadequate line-list for these molecules (e.g \citealt{Hoeijmakers2015}; \citealt{deRegt2022}), or a physical absence of the species at any detectable abundance in the planetary atmosphere. 


While such studies have been highly successful and have opened a new era in our ability to probe exoplanetary atmospheres, the cross-correlation techniques that are typically implemented suffer from a number of potential issues. For example, in the case of non-transiting planets (or where transits have not been recently observed), uncertainties in the planet's orbital period can compound over time, leading to errors in the orbital phasing of the planet. Due to the way that the cross-correlation function is calculated, minor errors in the orbital phase leads to smearing of any planetary signal, thereby weakening any detection \citep[as seen in][]{Guilluy2019, Webb2022}, while major errors can lead to the planetary signal being missed completely. In addition, cross-correlation approaches are subject to inherent aliasing issues due to contaminating spectral features that may arise due to overlapping lines and/or repeating line patterns. While some of these effects can be mitigated, such issues act to weaken any planetary signal. Recent developments in improving Doppler spectroscopy include attempts to combine low- and high-resolution observations \citep{Brogi2017}, and attempts to better retrieve atmospheric properties using log-likelihood methods \citep[e.g.][]{Gibson2020} as well as the implementation of  machine learning techniques \citep{Fisher2020}. Techniques such as these provide an added layer of improvement in the retrieval of planetary atmosphere properties. However, any new method that relies on cross-correlation will still suffer from some of the inherent underlying limitations of the technique.

In order to address this, \cite{Watson2019} suggested a new approach to extracting the planetary signal from high-resolution data-sets using Doppler tomography. This technique was initially developed to study and map the velocity flows in accreting binaries \citep{Marsh1988}. As applied to the study of exoplanet atmospheres, this technique has been shown to significantly reduce noise levels in simulated exoplanet detections, as well as reducing the effects of contaminating lines (e.g. incorrect line lists) and phase angle offsets (e.g. non-transiting planets). \cite{Watson2019} demonstrated that Doppler tomography was capable of recovering the signal of CO in the hot-Jupiter $\tau$ Bootis b as part of a pilot study to test its real-world application. However, the focus of \cite{Watson2019} was largely on tests using simulated data. The work in this paper represents the first full application of Doppler tomography to an exoplanet: HD 179949 b.

\subsection{HD 179949 b}
\label{sec:intro_planet}

HD 179949 b is a non-transiting planet that was discovered via radial velocity measurements by \cite{Tinney2001}. A list of parameters for the HD 179949 system can be found in Table \ref{tab:Params}. 

The first attempt at observing HD 179949 b at high resolution was performed by \cite{Barnes2008} who observed the planet using CRIRES, at a wavelength range of 2.1215-2.1740 $\micron$. Using deconvolution they were able to rule out the presence of absorption or emission lines due to H$_2$O, TiO and VO within their data. However, \cite{Rodler2013} have since noted that deconvolution  is less sensitive to molecular detection than the cross-correlation method. Analysis of this dataset may also have been inhibited by early analysis methods, which have been greatly improved in recent years \citep[e.g. by using SYSREM --][]{Cabot2019}.

The first successful measurement of the atmosphere of HD 179949 b was made by  \citet{Brogi2014} who detected the presence of carbon monoxide, as well as a weak signature of water on the day-side of the planet at $\sim$2.3 $\micron$. They were also able to determine the projected orbital velocity of HD 179949 b, which they found to be $142.8 \pm 3.4$ km s$^{-1}$. By directly determining the orbital velocity of a non-transiting planet in this way, both the orbital inclination and mass of the planet can be calculated. \cite{Brogi2014} found an orbital inclination of $i = 67.7 \pm 4.3$ degrees and a planetary mass of $M_P = 0.98 \pm 0.04$ $\mathrm{M_J}$ for HD 179949 b.

A more recent follow-up study by \cite{Webb2020} reported a marginal detection of water at $\sim$3.5 \micron. By combining the water signals seen in the 2.3 \micron~ data-set from \cite{Brogi2014} and their 3.5 \micron~data-sets they were able to further refine the planet's orbital velocity, obtaining a value of $K_p = 145.2 \pm 2.0$ km s$^{-1}$. While they were able to better constrain the planetary velocity, they note that further constraints on the planetary mass rely on improving the uncertainties associated with the stellar mass and the semi-major axis of the planet.

Each of the previous observations have focused on observations of the day-side of the planet, where the emission signal is expected to be at its strongest. However, as shown by \cite{deKok2014}, attempting to perform high-resolution Doppler spectroscopy on the cooler night side of the planet is also a viable option. They found that while the line profile will change depending on the $T$-$P$ profile of the day- and night-sides, the depth of these lines does not necessarily change. 


In this paper we  present the application of Doppler tomography to the planet HD 179949 b. We discuss our basic data reduction process in Section \ref{sec:data_reduction} and in Section \ref{sec:methods} we elaborate on the use of Doppler tomography. Within Section \ref{sec:results1} we present our Doppler tomography recoveries of both CO and H$_2$O signals from the atmosphere of HD 179949 b and directly compare this to results obtained using cross-correlation. Finally, we summarise our findings in Section \ref{sec:conclusions} and outline additional adaptations that may improve Doppler tomography recoveries of planetary atmosphere signals in the future.

\begin{table}
\centering
\caption{The stellar and planetary parameters for the HD 179949 system taken from the literature. a) \protect\cite{Takeda2007}, b) \protect\cite{Gray2006}, c) \protect\cite{Wittenmyer2007}, d) \protect\cite{Brogi2014}, e) \protect\cite{Butler2006}. For this paper we have assumed an eccentricity, $e$, of 0.}
 
\begin{tabular}{cccc}
\hline
\hline 
Parameter & & & Value 
\\ \hline
\textbf{HD 179949} & & &  \\
$M_{\star}$ (M$_{\odot})$ & & & 1.181 $\pm$  0.039~$^\mathrm{a}$\\
$R_{\star}$ (R$_{\odot})$ & & & 1.22 $\pm$ 0.05~$^\mathrm{a}$ \\
Spectral Type & & & F8 V ~$^\mathrm{b}$ \\
$T_{\mathrm{eff}}$ (K) & & & 6260 $\pm$ 44~$^\mathrm{c}$\\
$v_{\mathrm{sys}}$ (km s$^{-1})$ & & & -24.35 $\pm$ 0.18~$^\mathrm{c}$\\
$K_p$ (km s$^{-1})$ & &  & 142.8 $\pm$ 3.4~$^\mathrm{d}$

\\ \hline
\textbf{HD 179949 b} & & & \\
$T_{0}$ (JD) & & & 2455757.8190 $\pm$ 0.0022 ~$^\mathrm{d}$\\ 
$P$ (days) & & & 3.092514 $\pm$ 0.000032 ~$^\mathrm{e}$\\
$M_{p} \sin(i)$ (M$_\mathrm{J}$) & & & 0.92 $\pm$ 0.076 ~$^\mathrm{e}$\\

\end{tabular}
\label{tab:Params}
\end{table}

\section{Observations}
\label{sec:observations}

\begin{figure}
\centering

	\includegraphics[width = 8cm]{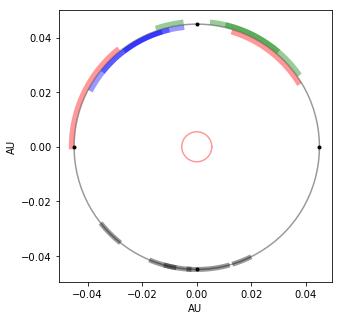}
    \caption{The orbital phases of HD 179949 b that are covered by the observations in our data set (see Table~\ref{tab:obs}). Pink represents the observations at 2.1 \micron, blue at 2.3 \micron\, and green at 3.5 \micron. The black represents the phase covered by the night-side 2.3 \micron\, observations.}
    \label{fig:phase_coverage}
\end{figure}

Our data-set consists of multiple archival observations of HD 179949 b, taken with the Cryogenic High Resolution Infrared Echelle Spectrograph (CRIRES; \citealt{Kaeufl2004}). These datasets are summarised in Table~\ref{tab:obs}. CRIRES was a high resolution (of up to R$\sim$100,000) spectrograph that was mounted at the Nasmyth-B focus of UT1 on the Very Large Telescope (VLT) in Cerro Paranal. CRIRES was composed of four Aladdin III detectors with 1024×512 pixels each. As is typical with CRIRES observations, the telescope was nodded along the slit in an ABBA pattern to improve the background subtraction. In total 1015 spectra were obtained over 13 observing nights spanning $\sim$7 years. The overall phase coverage is shown in Fig.~\ref{fig:phase_coverage}.

During each individual night the wavelength ranges were fixed, but a total of three different spectral ranges were observed. The data include both day-side and night-side observations. In this work we have analysed the day-side and night-side spectra separately, where the day-side spectra are defined as those where the orbital phase is between 0.3 and 0.7, and the night-side is between orbital phases of 0.8 and 0.2. We separate the day- and night-side observations in this manner since they are likely to exhibit different $T$-$P$ profiles. Furthermore, we also treat each of the 3 distinct wavelength ranges separately. 

\begin{table*}
\caption{A table showing the observation date of each of the datasets included within this study, alongside the wavelength range, phase coverage, resolution, and number of spectra taken for each night. The ESO program ID code is also included.}
\label{tab:obs}

\begin{tabular}{ccccccc}
\hline
\textbf{Observation Date} & \textbf{Spectral Range (\textmu m)} & \textbf{Resolution} & \textbf{Phase Coverage} & \textbf{Number of Spectra}  & \textbf{ESO ID} \\ 
\hline July 26 2007  & 2.12 - 2.17  & 50,000 & 0.34 - 0.45  & 94  & 079.C-0373  \\
August 02 2007  & 2.12 - 2.17  & 50,000 & 0.61 - 0.75  & 74  & 079.C-0373 \\
July 16 2011  & 2.28 - 2.35  & 100,000 & 0.55 - 0.67  & 197 & 186.C-0289 \\
July 19 2011  & 2.28 - 2.35  & 100,000 & 0.52 - 0.64  & 193  & 186.C-0289 \\
August 22 2011  & 2.28 - 2.35  & 100,000 & 0.54 - 0.60  & 106 & 186.C-0289 \\
April 18 2014  & 2.28 - 2.35  & 100,000 & 0.94 - 0.97  & 40  & 093.C-0676 \\
April 25 2014  & 3.45 - 3.54  & 100,000 & 0.52 - 0.55  & 40 & 093.C-0676 \\
April 28 2014  & 2.28 - 2.35  & 100,000 & 0.86 - 0.89  & 40 & 093.C-0676 \\

May 10 2014  & 2.28 - 2.35  & 100,000 & 0.05 - 0.07  & 44 & 093.C-0653 \\
June 8 2014  & 3.45 - 3.54  &100,000 & 0.39 - 0.46  & 78  & 093.C-0676 \\
June 23 2014  & 2.28 - 2.35  &100,000 & 0.95 - 0.99  & 48 & 093.C-0653 \\
July 12 2014  & 3.45 - 3.54  &100,000 & 0.35 - 0.38  & 40 & 093.C-0676 \\
July 14 2014  & 2.28 - 2.35  &100,000 & 0.99 - 0.04  & 71 & 093.C-0653 \\

\end{tabular}
\end{table*}

\section{Data Reduction}
\label{sec:data_reduction}

\begin{figure*}
\centering

	\includegraphics[width = 16cm]{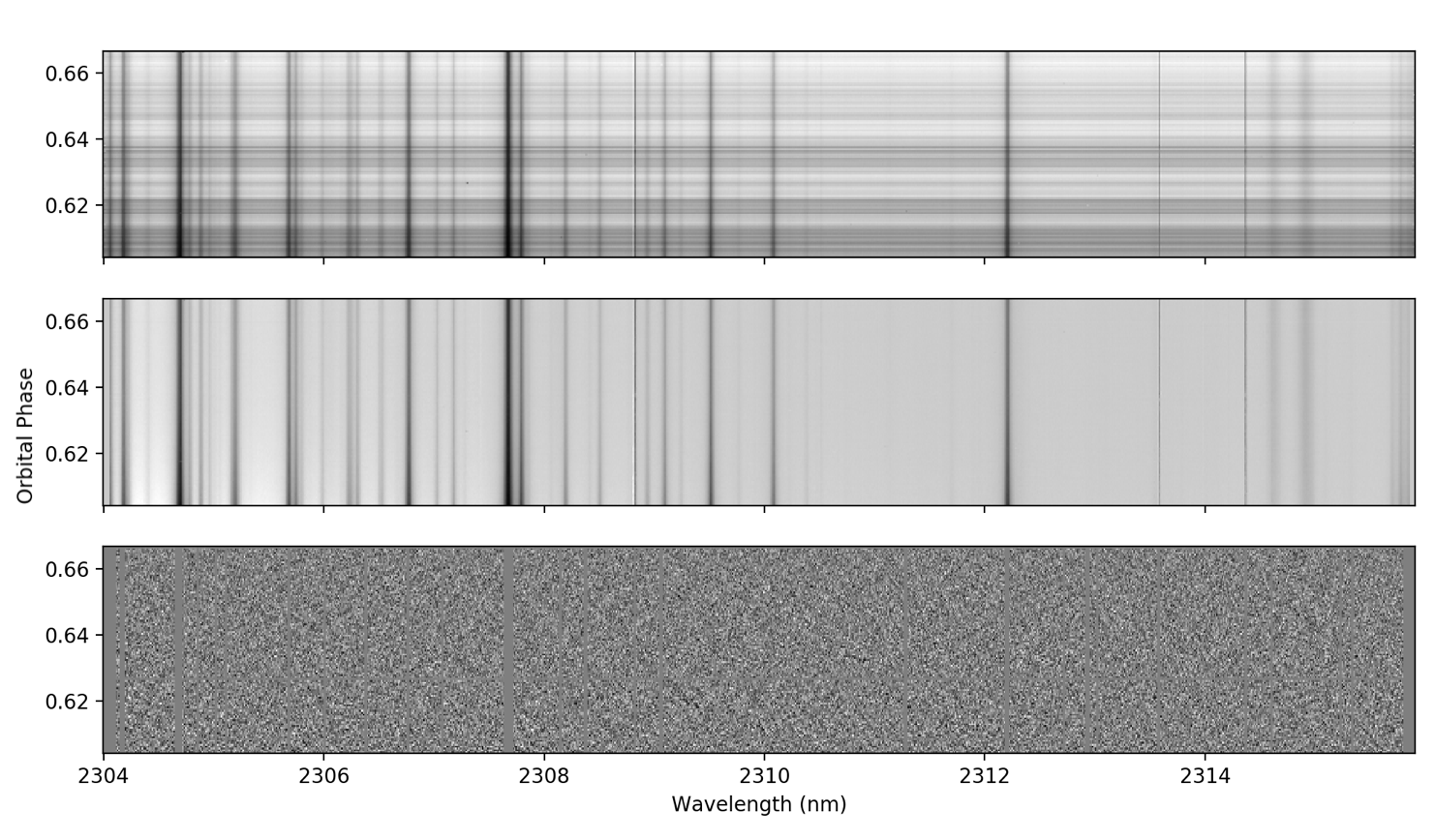}
    \caption{An example of the detrending process as applied to data taken on July 16 2011. Top panel: the data after extraction using the CRIRES pipeline and the correction of bad pixels. Middle panel: the same data after aligning and normalising the spectra. Bottom panel: the final de-trended data  after applying SYSREM, a high pass filter, and dividing each column by its standard deviation. Particularly noisy regions of the data have also been masked. It is the de-trended data presented in the bottom panel that is used in the subsequent Doppler tomography and cross-correlation analysis.
    }
    \label{fig:Detrending}
\end{figure*}

We extracted the spectra using the standard CRIRES pipeline (v2.3.2) through the \textsc{esorex} command line tool. This implements dark subtraction, flat-fielding, non-linearity and bad-pixel correction, as well as the combination of each AB nodding pair and the optimal extraction of each one-dimensional spectrum \citep[][]{Horne1986}. An example of the extracted spectra can be found in the first panel of Figure \ref{fig:Detrending}.

Subsequent data reduction steps follow standard practice  within the high resolution spectroscopy community (e.g \cite{Brogi2012}, \cite{Birkby2013}). For this, each individual night and detector were treated independently in the data reduction process. 

First, individual bad pixels that were not identified by the CRIRES pipeline were identified by eye and corrected via interpolation with their nearest neighbours, while larger sections containing many adjacent bad pixels were masked. Second, while an initial wavelength solution is provided by the pipeline, this is inadequate for our purposes due to the high precision required. We therefore selected the spectrum with the highest signal-to-noise ratio as a reference spectrum. The remaining spectra were then aligned to this reference spectrum. This was done by cross-correlating a set of prominent telluric lines within each spectrum to determine the shift, and applying this shift using a global spline interpolation. This resulted in small sub-pixel shifts, and allowed each spectrum to be placed on a common wavelength grid. To further improve the wavelength solution, we identified several strong, unblended telluric lines from a model made using ATRAN \citep{Lord1992}, and determined their exact positions in the data in order to determine an updated wavelength solution. The computed wavelength solution for each night had an average error of $\sim$ 0.10 -- 0.35 $\mathrm{km\,s^{-1}}$.

Subsequently, the spectra were  normalised by fitting a second order polynomial to the continuum. This was done by performing iterative sigma clipping, with a harsher sigma fitting for points lying below the polynomial, to clip out absorption lines and push the fit to the top of the continuum. Finally, a small number of pixels (typically 1-2$\%$ at the edges and in particularly noisy regions that exceeded two times the average variance) were masked. The aligned, normalised spectra can be seen in the second panel of Figure \ref{fig:Detrending}.
Finally, the orbital phasing of HD 179949 b was calculated from the period and time of conjunction given in Table~\ref{tab:Params} assuming a circular orbit. 



\subsection{Removing telluric features}
\label{sec:sysrem}

A vital component of high-resolution Doppler spectroscopy is the detrending process, where the stationary stellar and telluric features are removed from the spectrum, leaving only the Doppler shifted planetary lines and residual noise. In recent works, the detrending process has often relied on an algorithm known as SYSREM \citep{Tamuz2005}. This was first implemented by \cite{Birkby2013} and has been widely implemented since. The SYSREM algorithm involves principal component analysis (PCA), where subsequent iterations remove systematic trends, while also taking the errors of the data into account.

It can be difficult to determine what parameters SYSREM fits and removes. For example, in \cite{Birkby2013}, the first SYSREM iteration corresponds closely with the removal of the airmass trend. However, subsequent iterations do not seem to correspond to any other physical trends, such as pressure or seeing. While SYSREM requires no priors, there is debate over the optimal number of SYSREM iterations to perform on the data. The number of SYSREM iterations is often optimised by injecting an artificial planetary signal into the data. The optimal number of SYSREM iterations is that which gives this recovery the highest significance. We employ this method within our work, however, it should be noted that optimising recovery using SYSREM in this way can bias the results and lead to spurious signals \citep{Cabot2019}. On the other hand, applying too many SYSREM iterations risks over-fitting the systematic features and removing portions of the planetary signal resulting in a possible non-detection \citep{Birkby2017}. 

In order to ensure the validity of our detections we have also generated maps by calculating the optimal number of SYSREM iterations for each detector in a model-independent manner. By calculating the RMS (root mean square) of the data after each iteration, we are able to identify a point at which further iterations have little effect on the RMS value. Using this method, we are still able to recover a signal at the planetary position in each of our maps (see Section \ref{sec:appendix} in the Appendix for more details).

After applying SYSREM to our data, remaining low-order structure was removed by applying  both a second order polynomial and a Gaussian high-pass filter, with a standard deviation of 15 pixels, to each row. This reduces noise while preserving the correlated planetary signal. Wavelength bins with particularly strong residuals were masked. Each bin was then divided by its standard deviation. A final adjustment was made to the wavelength to correct for the barycentric velocity and the systemic velocity of the star-planet system so the data is in the stellar rest-frame. The final product of the detrending process can be seen in the third panel of Figure \ref{fig:Detrending}.

\section{Methods}
\label{sec:methods}
The planetary features that we are attempting to recover are hidden beneath the noise level of our data. To extract such signals, we need to employ techniques that combine the signals from many lines to effectively boost the signal-to-noise of any planetary detection.
For the analysis presented in this work we apply both conventional cross-correlation methods as well as Doppler tomography. In this section we briefly outline these methods as implemented here, as well as the generation of the relevant line-lists and planetary atmospheric models that are required by these methods.

\subsection{Cross-correlation}
\label{sec:CCF}
In order to compare the performance of Doppler tomography with standard techniques, we applied cross-correlation to the fully reduced data-sets. We performed the cross-correlation using the \textsc{NumPy} function \texttt{numpy.correlate} by correlating our detrended data with a model atmospheric spectrum (see Section \ref{sec:models}). As part of the cross-correlation process, the model spectra were interpolated onto the wavelength grid of the observed data for each velocity lag computed. This approach was undertaken to prevent interpolation errors being introduced to the observed data.

We integrate the cross-correlated data along radial velocity curves with $K_{p}$ values ranging from $-225$ km s$^{-1}$ to $+225$ km s$^{-1}$ and systemic velocities ($v_{\mathrm{sys}}$) from $-225$ km s$^{-1}$ to $+225$, both in steps of 1.5 km s$^{-1}$. These were then used to create a phase-folded CCF (cross-correlation function) map, an example of which is presented in the right panel of Figure~\ref{fig:Brogi_CO_ind}. 

This process was repeated for a variety of different atmospheric models with a range of parameters presented in Table~\ref{tab:pt} and described in more detail in Section~\ref{sec:models}. The CCF maps that are presented within this paper are from models with atmospheric parameters that resulted in the highest signal-to-noise ratio (SNR, see Section \ref{sec:uncertainies} for more details).

\subsection{Doppler tomography}
\label{sec:DT} 
In contrast to CCF analysis, Doppler tomography fits all of the observed data. It does so assuming that the velocity of any object in a time-series of spectra is given by:

\begin{equation} \label{eq:1}
v_r(\phi)=\gamma-v_x \cos(2\pi\phi) + v_y \sin(2\pi\phi)
\end{equation}

\noindent where $v_r$ is the radial velocity of the planet, $\gamma$ is the systemic velocity of the star-planet system, $\phi$ is the phase angle of the planet, and $v_x$ and $v_y$ are the radial velocity semi-amplitude of the planet in the $x$ and $y$ directions, respectively. Similar to the CCF, Doppler tomography integrates the data over all phases for a range of $v_x$ and $v_y$, resulting in 2D map of the strength of the signal in velocity space --- known as a Doppler map (see, for example, the left panel of Figure~\ref{fig:Brogi_CO_ind}). We note that  $v_x$ in the Doppler tomograms is not comparable to $v_{\mathrm{sys}}$ in the CCF maps. This is because we must fix $v_{\mathrm{sys}}$ in Doppler tomography, but effectively leave the phase offset free. If the phasing is correct, the planetary signal ought to lie on the $v_x$ = 0 line. Any phase offset, however, causes a rotation of the Doppler tomography map and leads to the planetary signal no longer lying on this line (see  \citealt{Watson2019}). This contrasts with the CCF method, where the phase offset is fixed, but $v_{\mathrm{sys}}$ is allowed to vary.

For planets on circular orbits,
\cite{Watson2019} showed that Doppler tomography could be applied to the retrieval of exoplanetary atmospheres, comparable to the traditional cross-correlation methods used in studies such as \cite{Snellen2010} and \cite{Brogi2012}. In a similar manner to cross-correlation, Doppler tomography is applied on fully detrended high-resolution data. Therefore, absorption lines from individual atomic and molecular species can be investigated.

For a full outline of the procedure as applied to exoplanets, we refer the reader to \cite{Watson2019}. However, the main benefits are two-fold. As noted earlier, the first factor is that Doppler tomography fits the data, which reduces aliasing effects that CCF analyses suffer from, particularly with semi-regularly repeating line-spacing (as can occur in molecular features). The second factor is the inclusion of a regularisation statistic within the code (we implement the maximum entropy method) that acts to suppress any non-coherent noise signals. Both of these aspects help reduce `background' noise and artefacts in the generated Doppler maps relative to CCF maps, thereby allowing fainter planetary signals to be recovered.

The final outcome of the Doppler map is controlled by a user-defined aim $\chi^2$ value, which generates a map with a given entropy value. In order to calculate the entropy of the resulting Doppler map, we define a `default map'. This default map can contain a priori information on the expected location of any planetary signal, and also contains the values that Doppler map defaults to in the absence of any constraining data. For our purposes, we assume a uniform default map (i.e. we assume no a priori information on the position of any potential signal/s).

For the Doppler maps presented in this paper, we adopt a 300 by 300 pixel velocity grid in $v_{x}$ and $v_{y}$, with a 1.5 km s$^{-1}$ step-size. This covers velocities ranging from $-225$ km s$^{-1}$ to $+225$ km s$^{-1}$ in both directions.

Doppler tomography requires information regarding the expected line positions and relative strengths for each molecule being probed in the planetary atmosphere. How these line positions are obtained is detailed in Section~\ref{sec:models}.


Doppler tomography proceeds by iterating to the aim $\chi^2$, the selection of which is somewhat subjective. Fitting to too low a $\chi^2$ results in a featureless Doppler map, whereas too high a $\chi^2$ results in a map dominated by noise. Both of these situations often result in an iterative process that fails to converge. Therefore, while there is no well-defined means to predetermine what aim $\chi^2$ should be sought, both the above scenarios are easily recognised. Within this study, we have chosen the $\chi^2$ that maximises the SNR of the Doppler map (see Section \ref{sec:uncertainies} for details on the SNR calculation).

The final $\chi^2$ and entropy values, alongside the number of iterations required to converge to a solution for each of the Doppler maps produced within this work can be seen in Table \ref{tab:crires_entropy}.

\begin{table*}
\centering
\caption{Reduced $\chi^2$ and entropy values reached in the Doppler tomograms, as well as the number of iterations required for each molecular detection that has been reported within this work.}
\label{tab:crires_entropy} 
\begin{tabular}{cccccccc}
\hline
Molecule  & Wavelength ($\mu$m)  & $\chi^{2}$ & Entropy & Iterations  \\ 
\hline 
CO (Dayside) & 2.3  & 0.774469 & -0.642020  & 204 \\
H$_2$O & 2.3 & 0.774491 & -0.526826 & 69 \\
H$_2$O & 3.5 & 0.765760 & -0.910062 & 85 \\
H$_2$O & 2.1 & 0.735534 &  -0.802095  & 17 \\
CO (Nightside) & 2.3 & 0.777998  & -0.527438 & 8  \\
\end{tabular}
\end{table*}

\subsection{Uncertainty and significance estimation}
\label{sec:uncertainies}
When performing statistical analysis on CCFs, it is often assumed that the background noise is Gaussian in nature. While the noise is not strictly Gaussian, assuming it is close to Gaussian makes assigning a statistical significance exponentially more straightforward and is standard in the literature. One method of assigning significance is to take the peak value of the signal with the CCF and divide this by the standard deviation  of the noise, giving a signal-to-noise value for the detection \citep[e.g.][]{Cabot2019, Merritt2021}.

Another method of calculating a statistical significance is performed by using the Welch's t-test \citep{Welch1947}. This t-test is a statistical analysis to determine the probability that the means of two samples are drawn from the same parent distribution. The scipy.stats package can be used to calculate the t-test p-value for the two samples. This is performed on the directly cross-correlated values. By knowing the planetary velocity, it is possible to calculate the pixels that fall within the planetary trail. By comparing the distribution of these values with those that are `out-of-trail' (i.e. noise), we can calculate a statistical significance between the planetary `in-trail' and `out-of-trail' values \citep[e.g][]{Brogi2012, Webb2020}.

Unfortunately, a similarly straightforward method of calculating the significance of the Doppler tomography output is not available. This is due to the method used in the noise reduction within the Doppler tomograms. By utilising the maximum entropy method in our Doppler tomography analysis, the background noise within any Doppler map is non-Gaussian and therefore cannot be treated in the same manner as a CCF output. If we do perform a signal-to-noise calculation on a Doppler tomogram signal, we get highly inflated, nonphysical values that are highly dependent on the chosen $\chi^2$ and resultant entropy values.

It was therefore imperative that we develop a method of performing statistical validation on our Doppler tomography outputs. We do this by developing a method that is analogous to performing the Welch's t-test on cross-correlated data. Each pixel in the Doppler tomogram represents an intensity for each $v_{x}$ and $v_{y}$ value. The full-width half maximum of the line profile is set within the default map, in our case 1.5 km s$^{-1}$ and this allows us to back-project the Doppler tomogram intensities in a manner that resembles the intermediary cross-correlated values. This gives rise to a planetary trail, from which we are directly able to determine the `in-trail' and `out-of-trail' pixels and calculate a statistical significance from their distribution. While we can naively attempt to perform the Welch's t-test analysis on our Doppler tomography dataset, we find that for cases with little background noise, it is often the case that the two distributions are too distinct to allow for a valid calculation to be made. In addition, and most importantly, the distributions produced from the Doppler tomography trails are clearly non-Gaussian, violating the main hypothesis of the Welch's t-test. Therefore, we cannot quote an analogous value here. We instead determined a significance by calculating the 16-84\% ranges for both the `in-trail' and `out-of-trail' distribution and have quoted these values for both Doppler tomography and CCF (see CCF Projection and DT Projection in Tables \ref{tab:results_CCF} and \ref{tab:results_DT}, respectively, for this comparison).

Each of the planetary trails generated from the cross-correlation analysis and Doppler tomography projection can be seen in the Appendix.

We have found that, in general, the aim $\chi^2$ that produces the highest SNR of the planetary signal is very similar to the aim $\chi^2$ that produces the highest significance when calculating the 16-84\% ranges for both in-trail and out-of-trail distribution. Therefore, to save on computational time, we have presented the Doppler map that gives the highest SNR within this work.

\subsection{Generation of line-lists and atmospheric models}
\label{sec:models}

Cross-correlation and Doppler tomography analyses have different requirements for the extraction of a planetary signal from our dataset. For the CCF method, we cross-correlate the observed data with an atmospheric model. On the other  hand, Doppler tomography requires a list of line positions and (relative) strengths.

In order to generate our atmospheric emission models we used the python based radiative transfer code petitRadTrans \citep{Molliere2019}. We assumed a mean molecular weight of 2.3 and set the planetary radius and surface gravity as 1.23 $\mathrm{R_J}$ and log $g$ = 3.35, respectively.

We tested for a range of temperature-pressure profiles and molecular abundances following the method outlined by \cite{Brogi2014} and \cite{Webb2020}. This involved parameterising the $T$-$P$ profile. We define two points within the atmosphere of the planet, one at a lower pressure ($P_1$, $T_1$) and the other at a higher pressure ($P_2$, $T_2$). We assume that for pressures lower than $P_1$, the atmospheric temperature is isothermal at $T_1$. The same is true for pressures higher than $P_2$, with a temperature $T_2$.

For pressures in between $P_1$ and $P_2$, the temperature gradient is defined as:

\begin{equation} \label{eq:2}
\Delta T_{\mathrm{slope}} = \frac{T_{1} - T_{2}}{\log(P_1) - \log(P_2)}
\end{equation}

The values that we have adopted for $P_1, T_1, P_2$ and $T_2$ for the dayside analysis have been taken from \cite{Brogi2014} and are summarised in Table~\ref{tab:pt}. The volume mixing ratio (VMR) of each molecular species was also varied in accordance with the range also studied by \cite{Brogi2014}. Since the $T$-$P$ profile of the planetary night-side may vary significantly from the day-side, we have expanded this analysis for the night-side atmospheric models. The values of $P_1, T_1, P_2$ and $T_2$ in this case can be seen in Table \ref{tab:pt}.

The atmospheric model was converted into the line list format required by Doppler tomography by first removing the continuum by fitting a 2nd-order polynomial, and then fitting to the peaks in the model using scipy.signal.find\textunderscore peaks and normalising to their relative strength.

\begin{table*}
\caption{A list of the atmospheric parameters that we used to calculate the planetary models for the day-side (top) and night-side (bottom) detections of HD 179949 b. }
\label{tab:pt}
\begin{tabular}{cccccc}
\hline
& log10($P_1$) (bar)  & $T_1$ (K)  & log10($P_2$) (bar) & $T_2$ (K) & log(VMR)\\ 
\hline Day & -4.5, -3.5, -2.5, -1.5  & 1200, 1450, 2150  & 1  & 1950 & -8.5, -8.0 ... -1.5, -1.0  \\

\hline Night & -4.5, -3.5, -2.5, -1.5  & 500, 800, 1200, 1400, 1600  & 1  & 500, 800, 1200, 1400, 1600 & -8.5, -8.0 ... -1.5, -1.0  \\
\end{tabular}
\end{table*}

\section{Results and Discussion}
\label{sec:results1}

In this section we present the results of our analysis of HD 179949 b using both Doppler tomography and cross-correlation. We have searched for signatures of H$_2$O on both the day- and night-side separately, considering each of the 3 different wavelength ranges covered by our data in turn. Analysing each wavelength range individually allows us to potentially probe different atmospheric layers as well as test the veracity of any H$_2$O detection by sampling different telluric contamination contributions. For CO, we only consider the 2.3\micron\, wavelength range where the CO band is observable, but investigate the planetary day- and night-sides independently. Finally, we search for other molecular features, such as HCN and CH$_4$ within each of the datasets.

\begin{figure*}
    \subfloat{%
        \includegraphics[width=.48\linewidth]{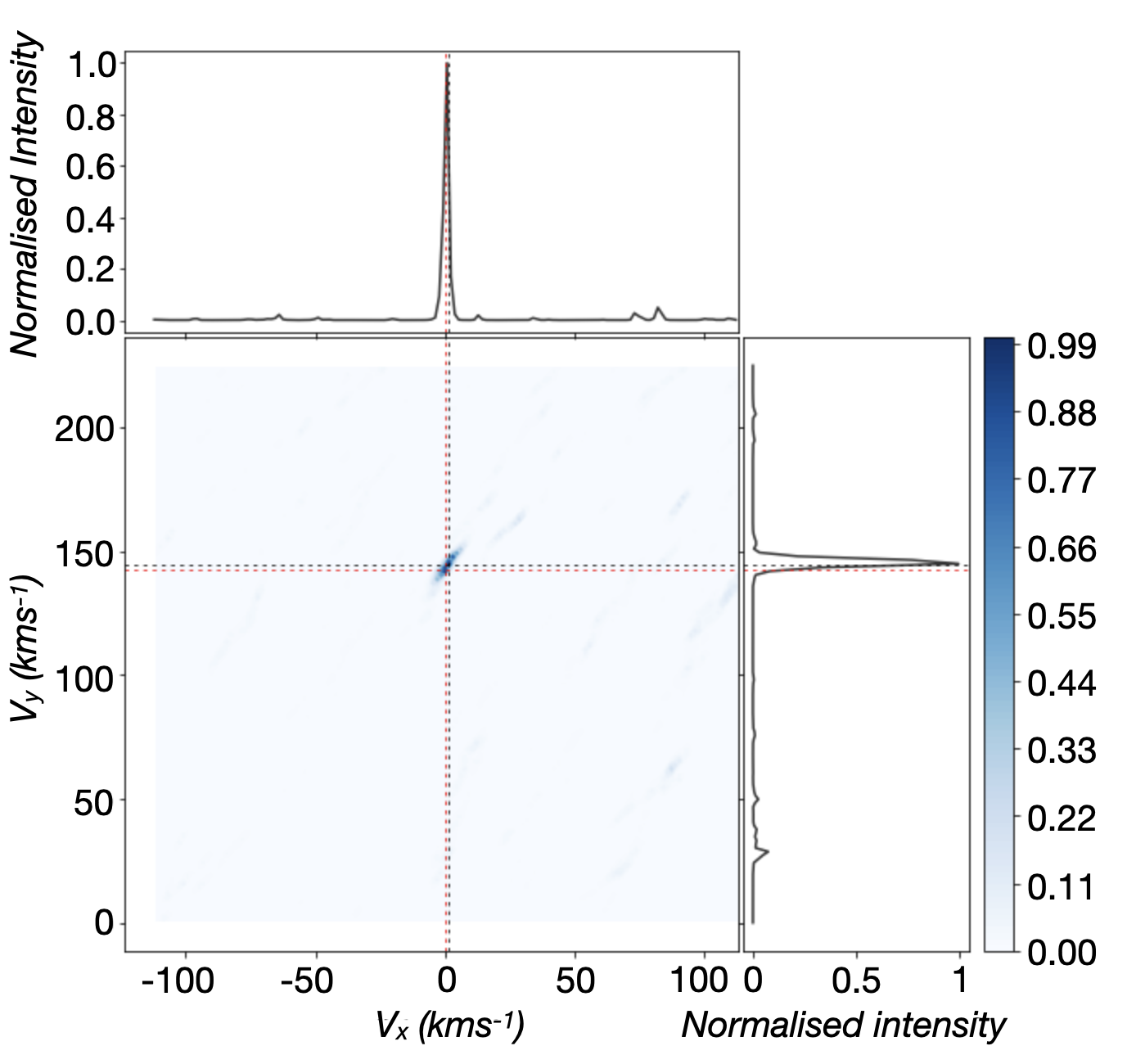}%
        \label{subfig:1a}%
    }\hfill
    \subfloat{%
        \includegraphics[width=.48\linewidth]{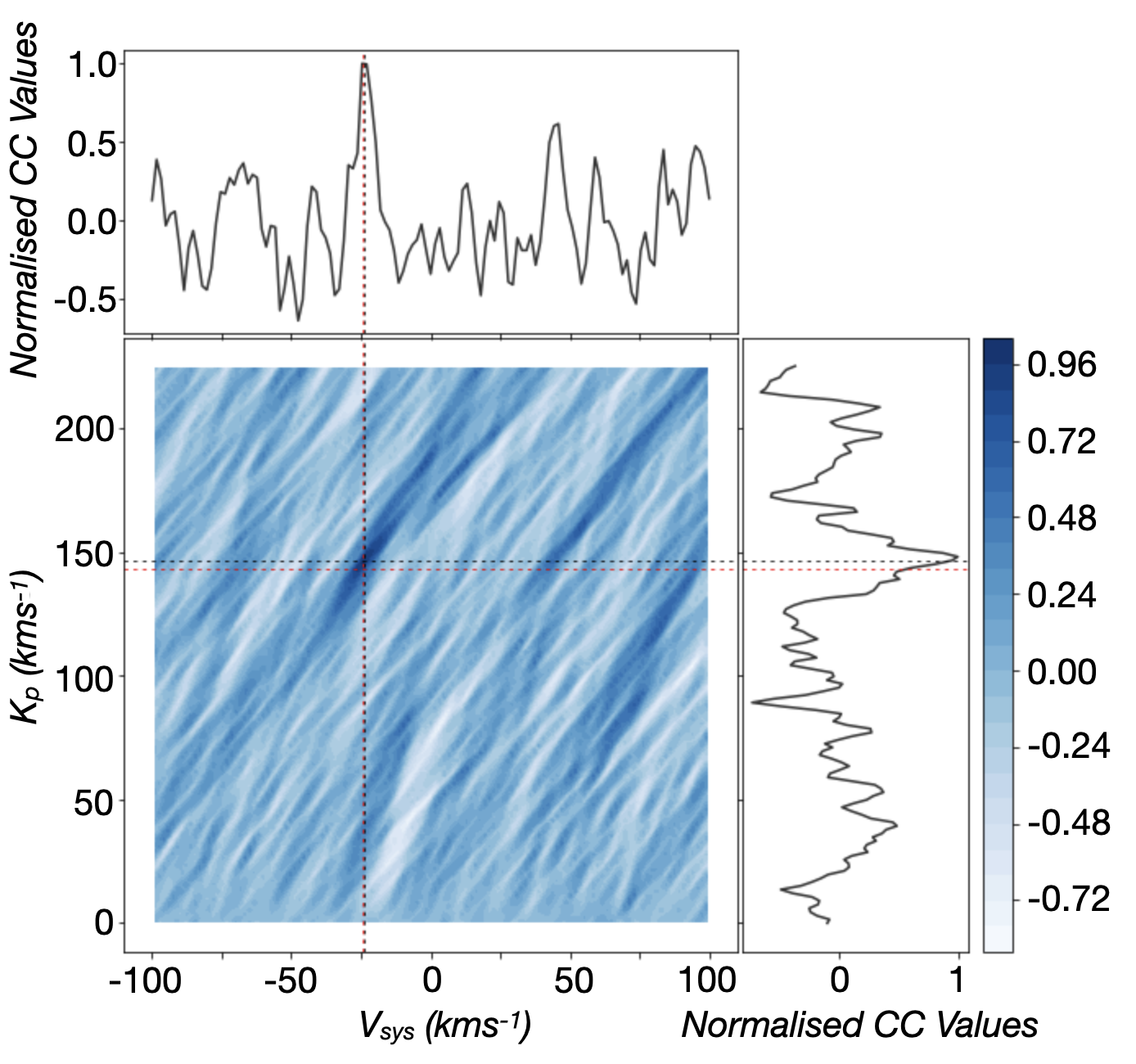}%
        \label{subfig:1b}%
    }\\
    \subfloat{%
        \includegraphics[width=.48\linewidth]{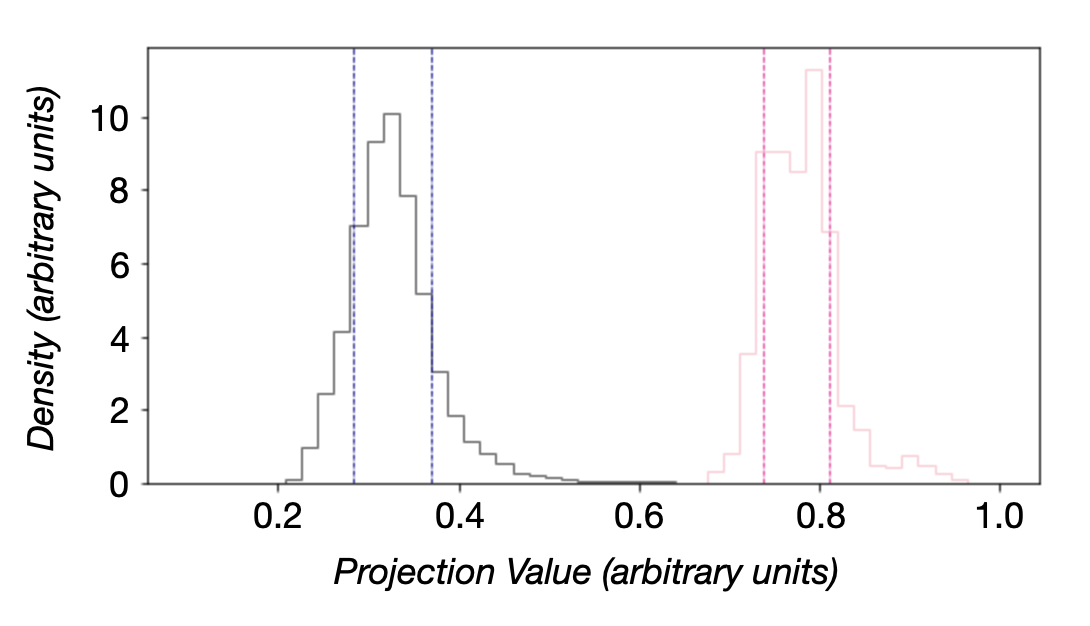}%
        \label{subfig:1c}%
    }\hfill
    \subfloat{%
        \includegraphics[width=.48\linewidth]{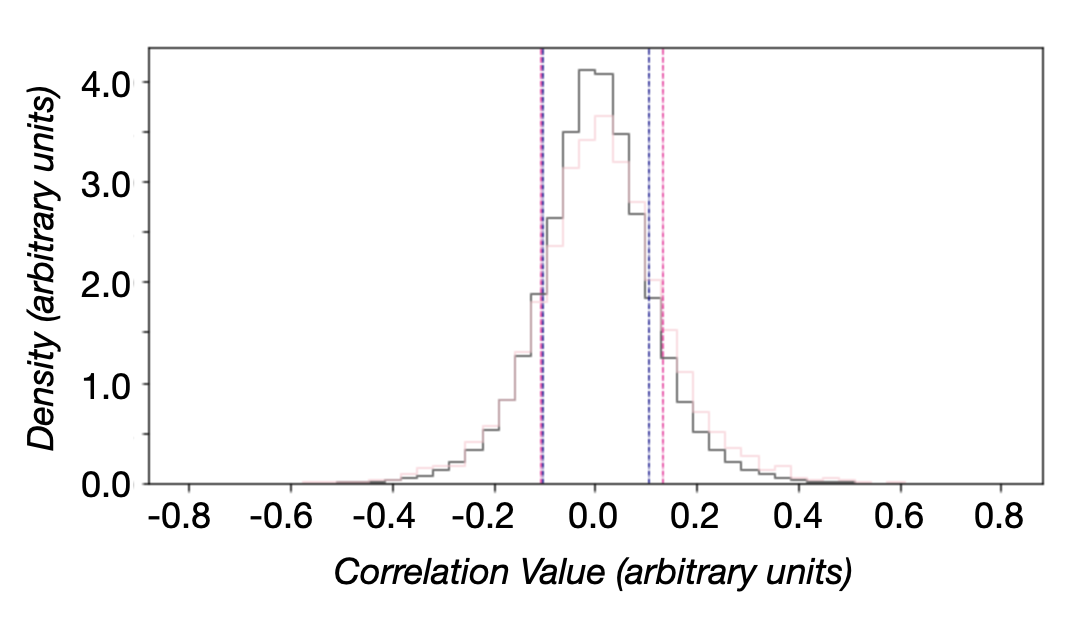}%
        \label{subfig:1d}%
    }
    
    \caption{A comparison of Doppler tomography (top left) versus cross-correlation
    (top right) for the recovery of the carbon monoxide signal at 2.3 \micron, including
    cross-sections through $v_x$ = 0 km s$^{-1}$ and $v_{\mathrm{sys}}$ in the Doppler
    tomogram and CCF map, respectively, as well as the peak planetary velocity. Here the black dashed lines
    indicate the position of the peak velocity (see Table \ref{tab:Kps}) and the red dashed
    lines indicate the expected planetary velocity given by Brogi et al. (2014),
    ($K_p = 142.8$ km s$^{-1}$). The bottom panels show the difference in distribution of
    the `in-trail' (pink) and `out-of-trail' (black) signals for each method. Vertical
    lines indicate the 16-84 percentiles.}

    \label{fig:Brogi_CO_ind}
    \end{figure*}

\begin{figure*}
    \subfloat{%
        \includegraphics[width=.48\linewidth]{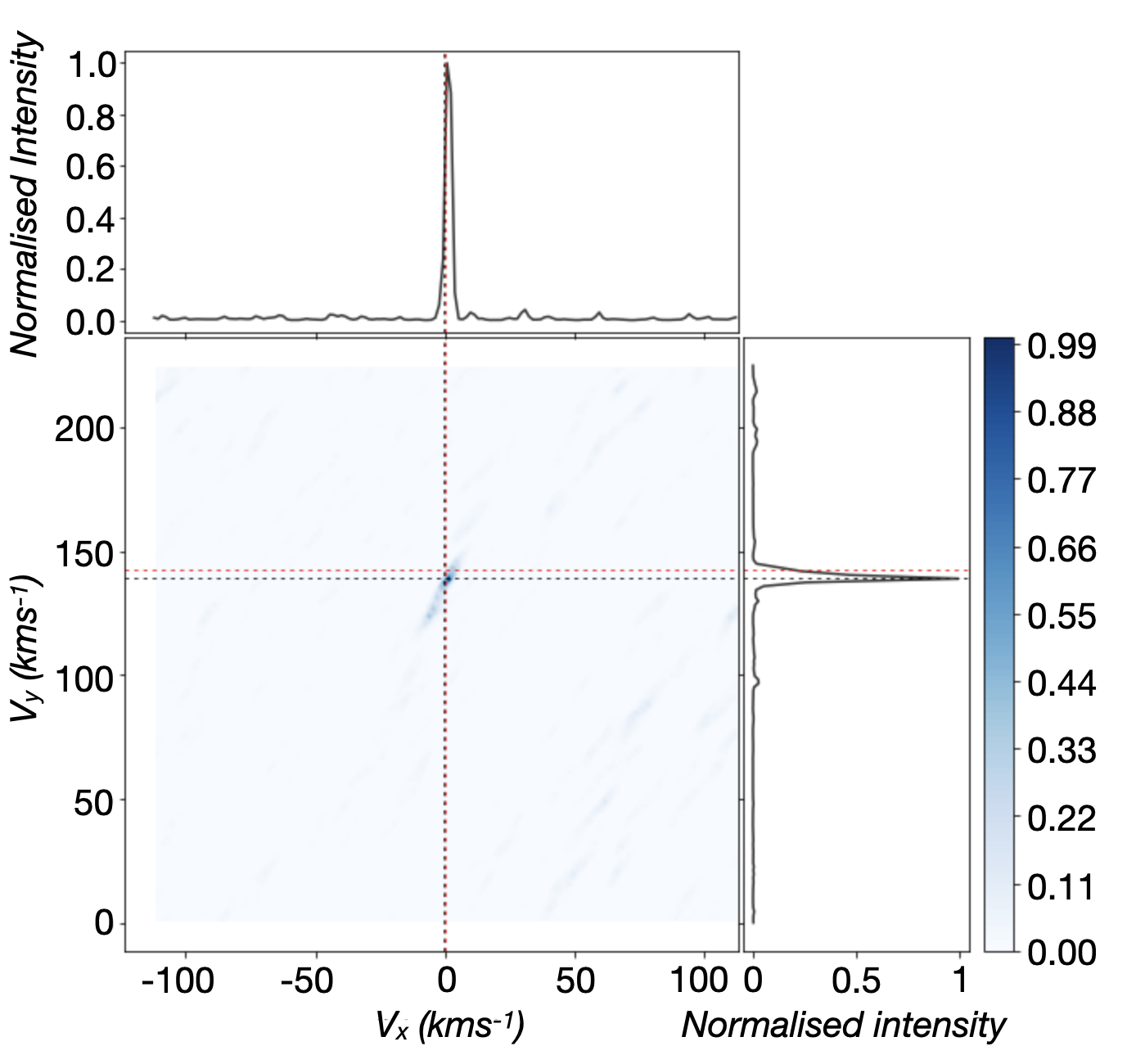}%
        \label{subfig:2a}%
    }\hfill
    \subfloat{%
        \includegraphics[width=.48\linewidth]{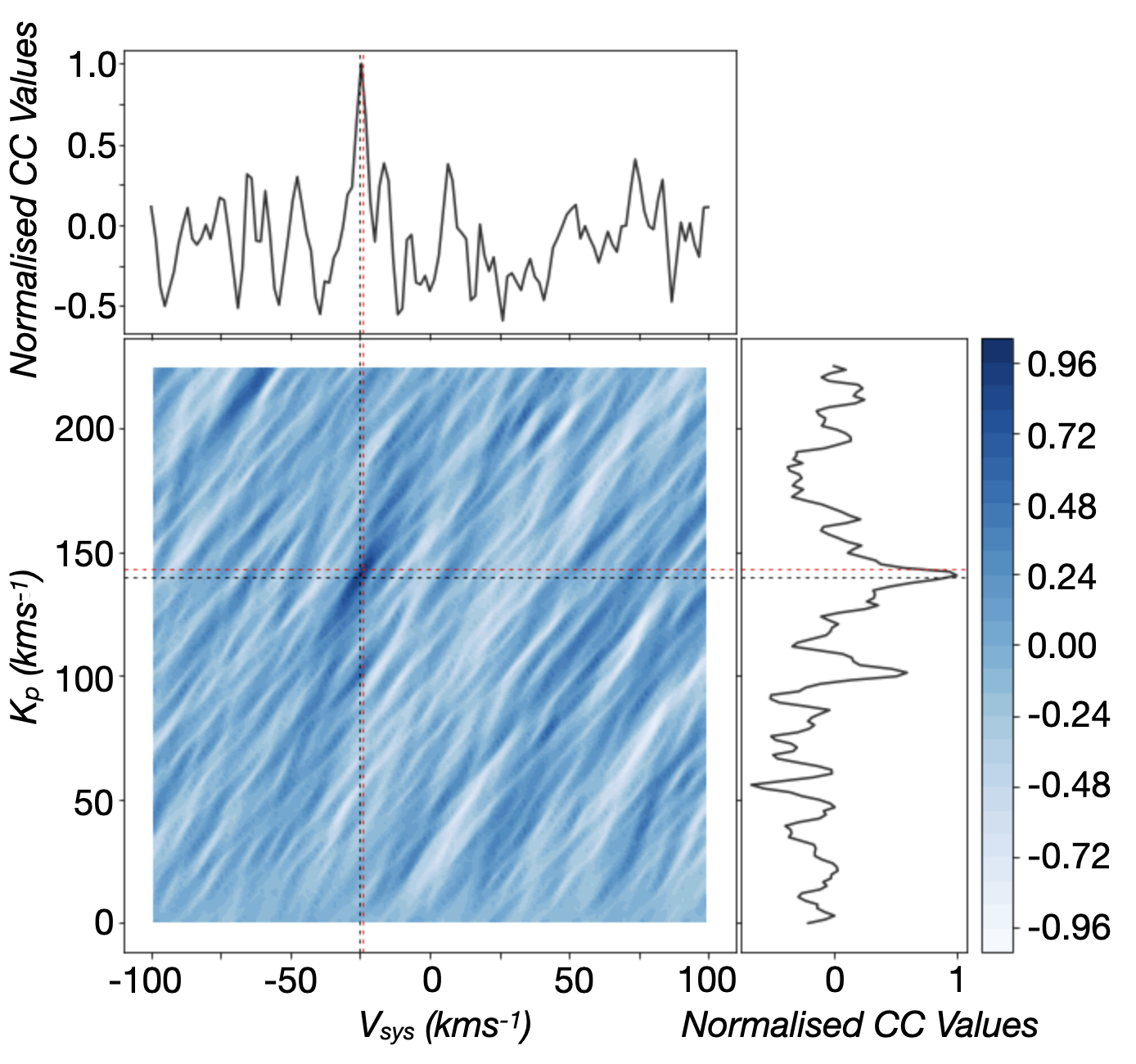}%
        \label{subfig:2b}%
    }\\
    \subfloat{
        \includegraphics[width=.48\linewidth]{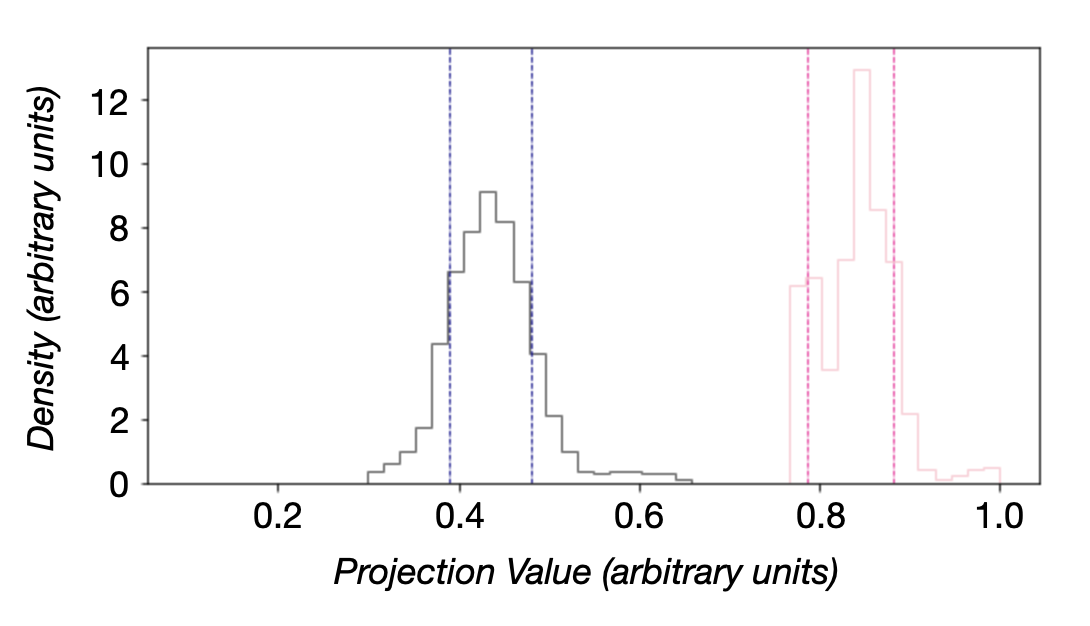}%
        \label{subfig:2c}%
    }\hfill
    \subfloat{%
        \includegraphics[width=.48\linewidth]{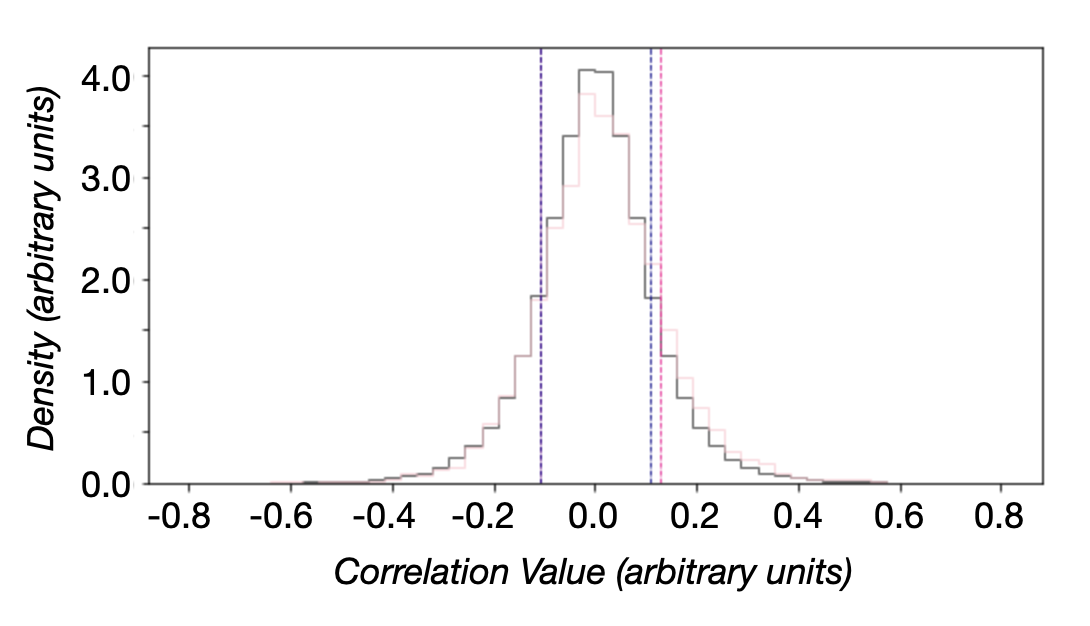}%
        \label{subfig:2d}%
    }
    \caption{A comparison of Doppler tomography (top left) versus cross-correlation (top right) for the recovery of the water signal at 2.3 \micron, including cross-sections through $v_x$ = 0 km s$^{-1}$ and $v_{\mathrm{sys}}$ in the Doppler tomogram and CCF map, respectively, as well as the peak planetary velocity. Here the black dashed lines indicate the position of the peak velocity (see Table \ref{tab:Kps}) and the red dashed lines indicate the expected planetary velocity given by Brogi et al. (2014), ($K_p = 142.8$ km s$^{-1}$).  The bottom panels show the difference in distribution of the `in-trail' (pink) and `out-of-trail' (black) signals for each method. Vertical lines indicate the 16-84 percentiles.}

    \label{fig:Brogi_H2O_ind}
    \end{figure*}

\begin{figure*}
    \subfloat{%
        \includegraphics[width=.48\linewidth]{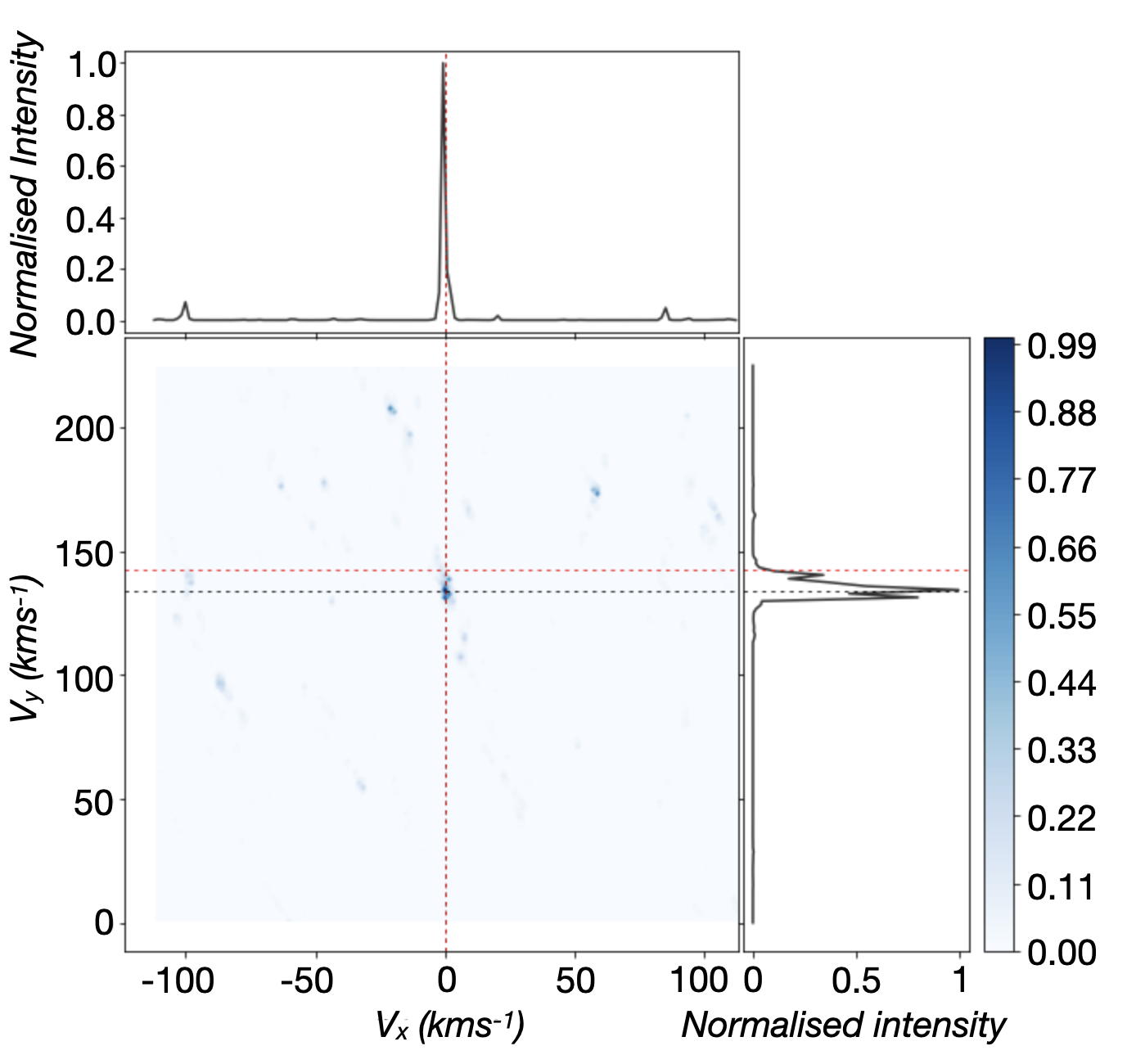}%
        \label{subfig:3a}%
    }\hfill
    \subfloat{%
        \includegraphics[width=.48\linewidth]{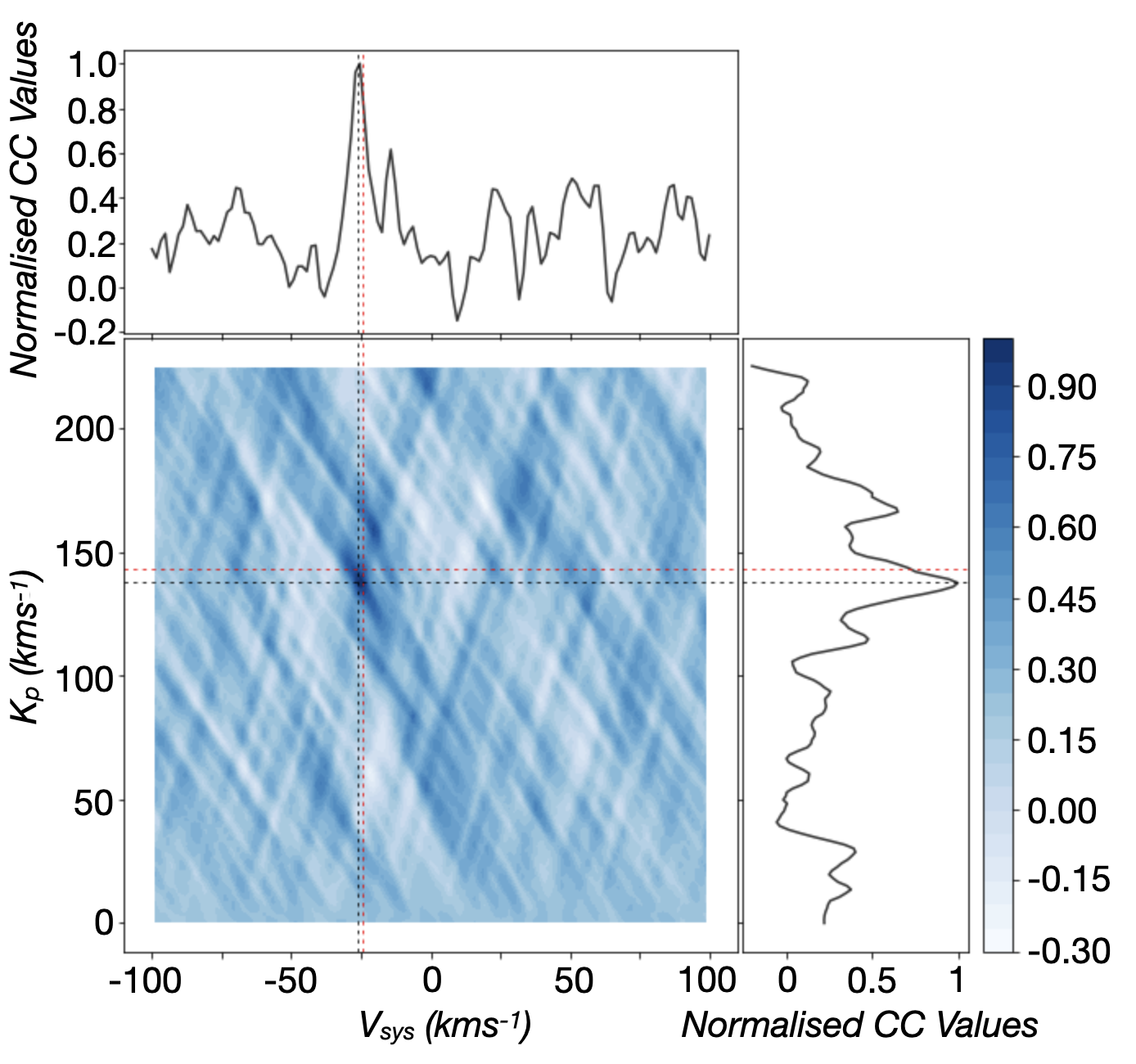}%
        \label{subfig:3b}%
    }\\
    \subfloat{
        \includegraphics[width=.48\linewidth]{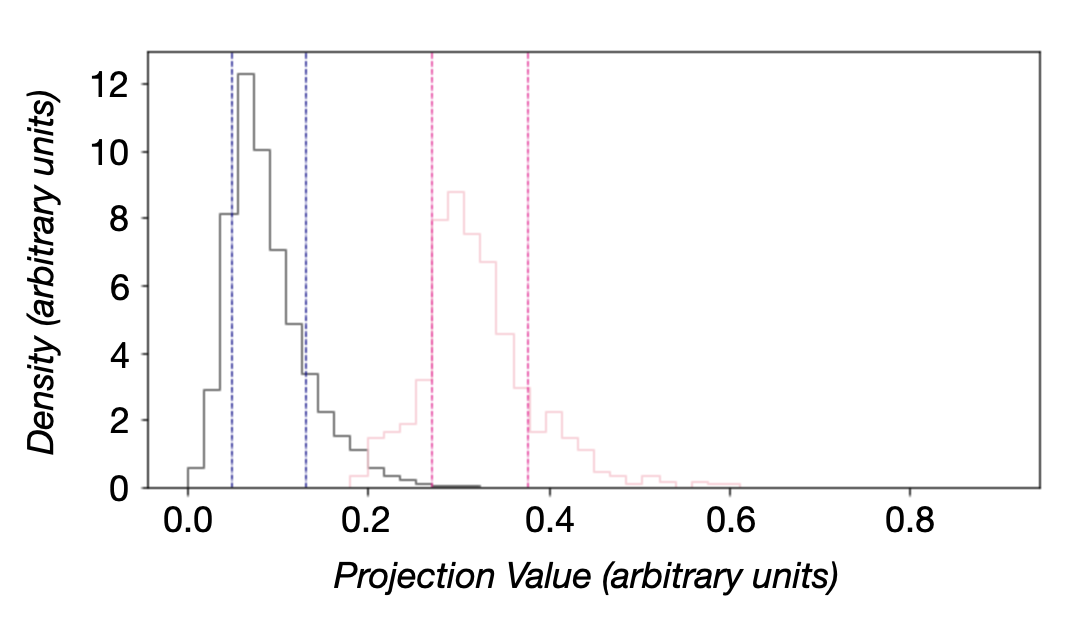}%
        \label{subfig:3c}%
    }\hfill
    \subfloat{%
        \includegraphics[width=.48\linewidth]{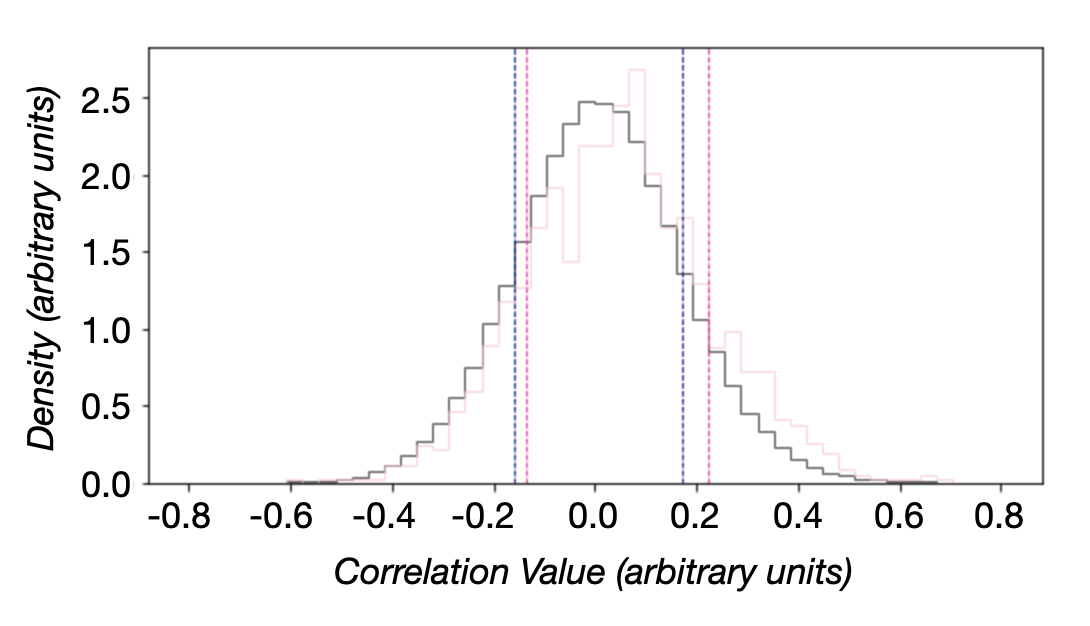}%
        \label{subfig:3d}%
    }
    \caption{A comparison of Doppler tomography (top left) versus cross-correlation (top right) for the recovery of the water signal at 3.5\,\micron, including cross-sections through $v_x$ = 0 km s$^{-1}$ and $v_{\mathrm{sys}}$ in the Doppler tomogram and CCF map, respectively, as well as the peak planetary velocity. Here the black dashed lines indicate the position of the peak velocity (see Table \ref{tab:Kps}) and the red dashed lines indicate the expected planetary velocity given by Brogi et al. (2014), ($K_p = 142.8$ km s$^{-1}$). The bottom panels show the difference in distribution of the `in-trail' (pink) and `out-of-trail' (black) signals for each method. Vertical lines indicate the 16-84 percentiles.}

    \label{fig:H2O_35}
    \end{figure*}

\begin{figure*}
    \subfloat{%
        \includegraphics[width=.48\linewidth]{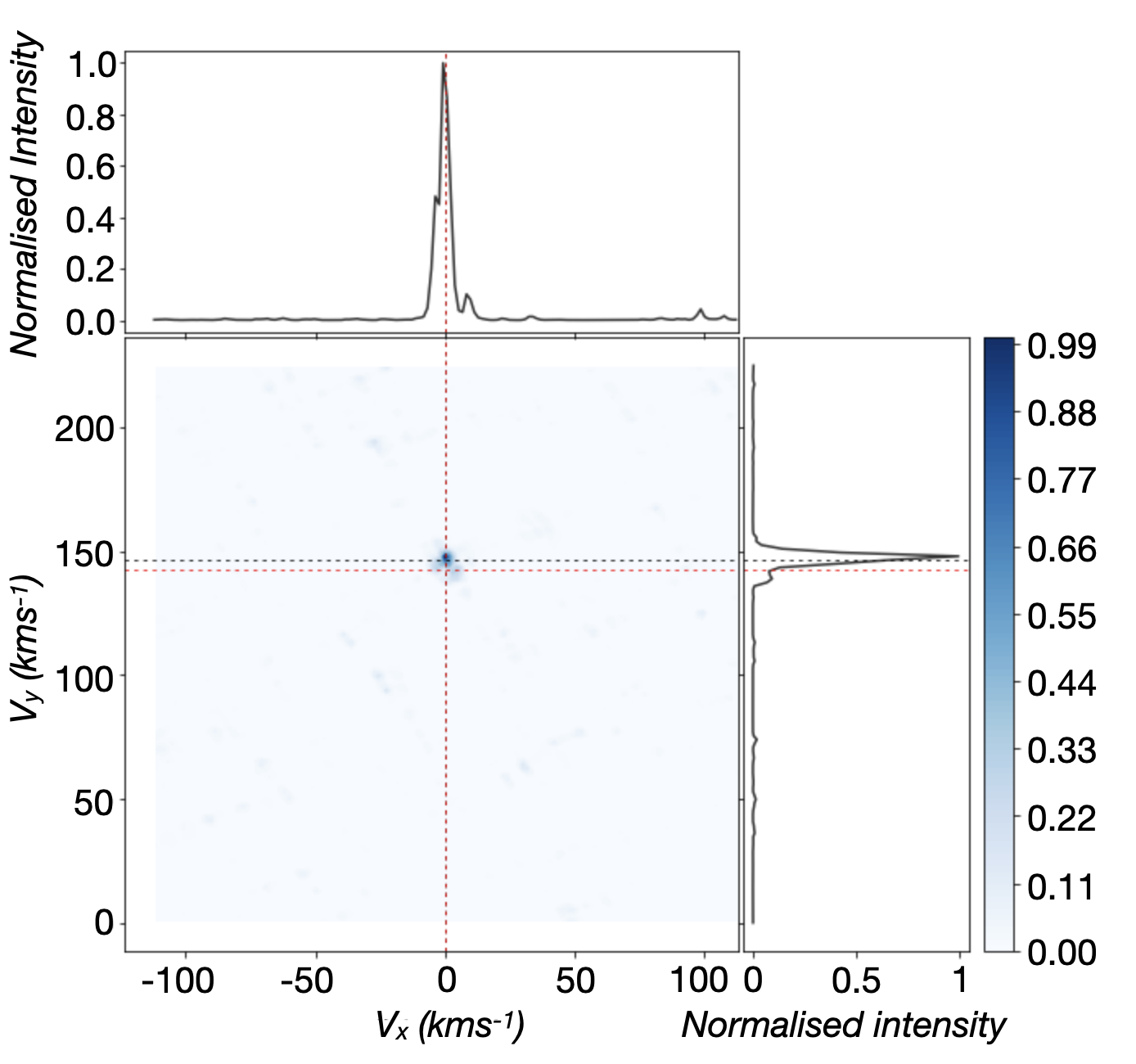}%
        \label{subfig:4a}%
    }\hfill
    \subfloat{%
        \includegraphics[width=.48\linewidth]{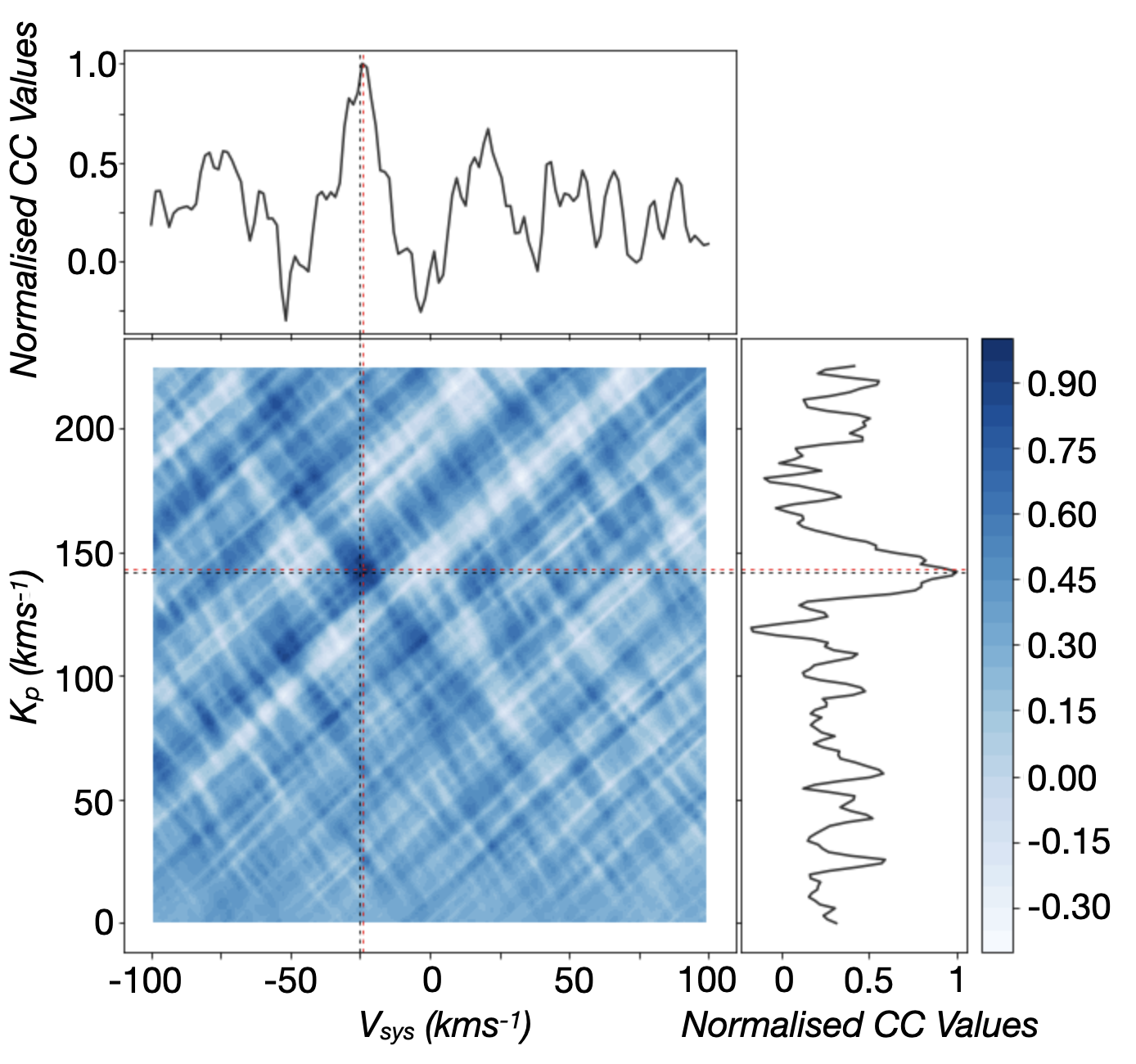}%
        \label{subfig:4b}%
    }\\
    \subfloat{
        \includegraphics[width=.48\linewidth]{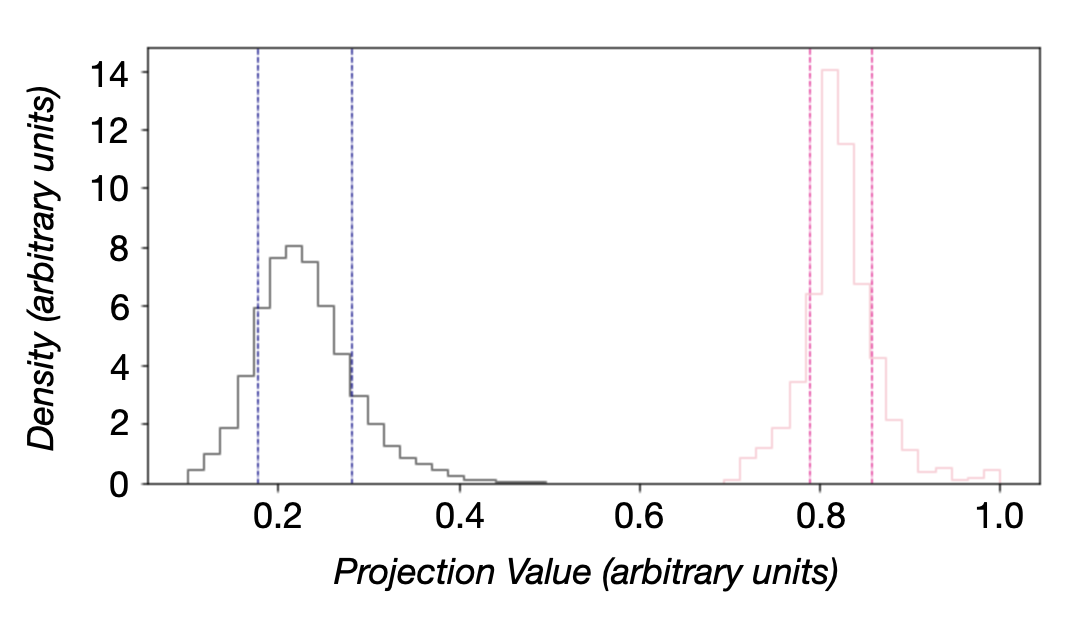}%
        \label{subfig:4c}%
    }\hfill
    \subfloat{%
        \includegraphics[width=.48\linewidth]{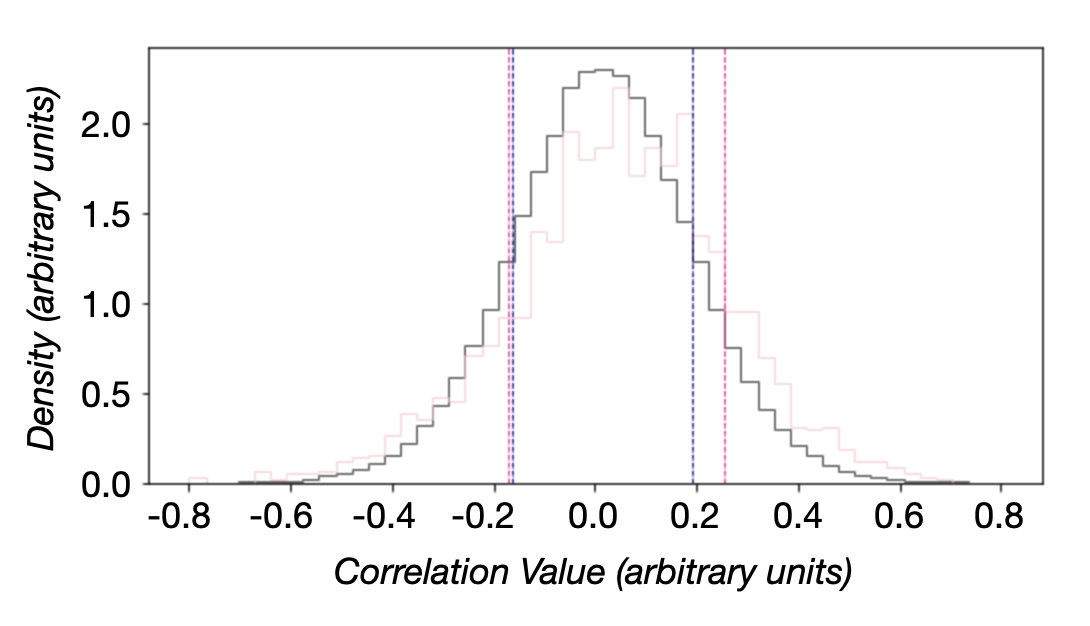}%
        \label{subfig:4d}%
    }
    \caption{A comparison of Doppler tomography (left) versus cross-correlation (right) for the recovery of the water signal at 2.1 \micron, including cross-sections through $v_x$ = 0 km s$^{-1}$ and $v_{\mathrm{sys}}$ in the Doppler tomogram and CCF map, respectively, as well as the peak planetary velocity. Here the black dashed lines indicate the position of the peak velocity (see Table \ref{tab:Kps}) and the red dashed lines indicate the expected planetary velocity given by Brogi et al. (2014), ($K_p = 142.8$ km s$^{-1}$). The bottom panels show the difference in distribution of the `in-trail' (pink) and `out-of-trail' (black) signals for each method. Vertical lines indicate the 16-84 percentiles.}

    \label{fig:H2O_21}
    \end{figure*}

\begin{figure*}
    \subfloat{%
        \includegraphics[width=.48\linewidth]{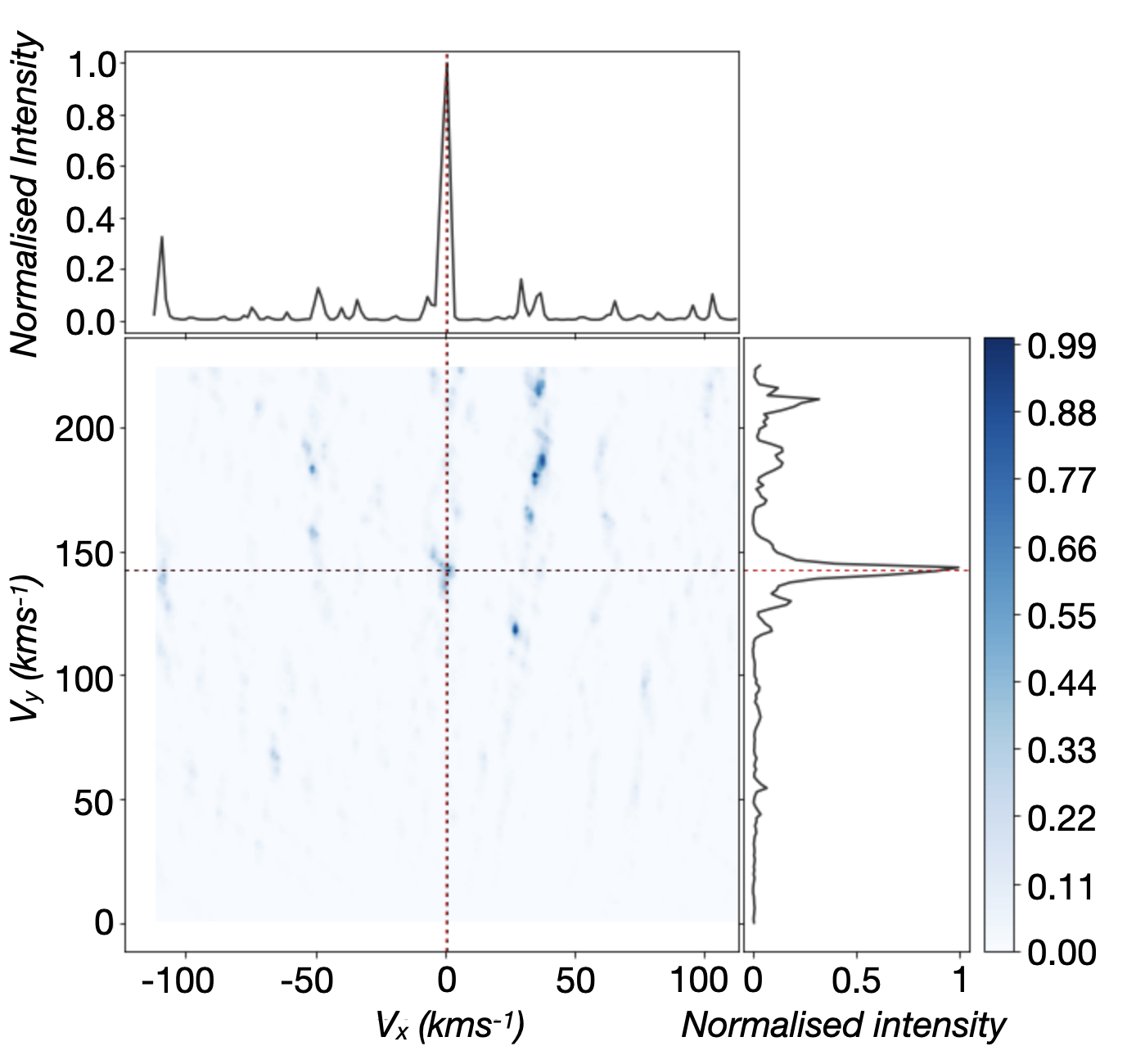}%
        \label{subfig:5a}%
    }\hfill
    \subfloat{%
        \includegraphics[width=.48\linewidth]{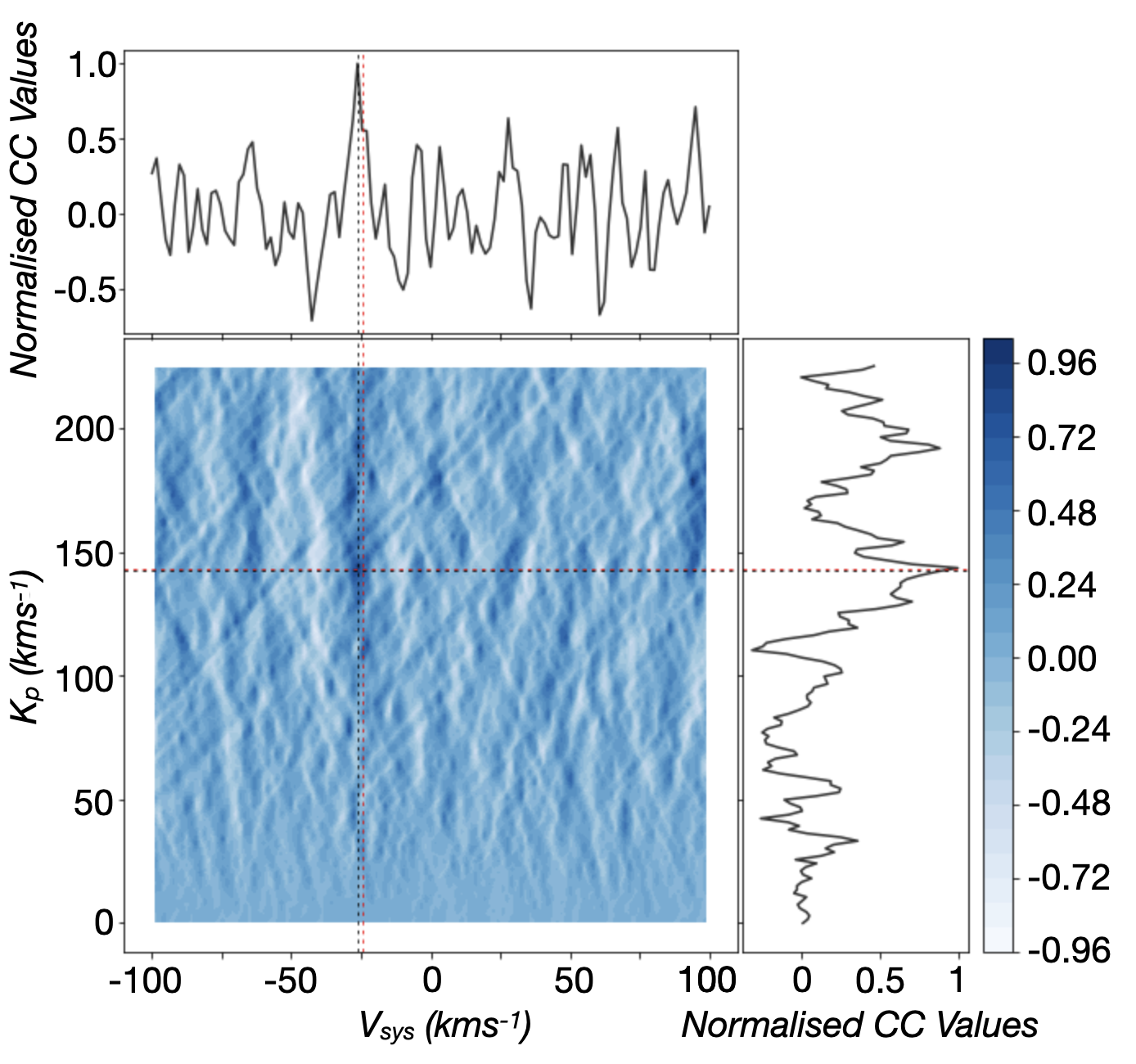}%
        \label{subfig:5b}%
    }\\
    \subfloat{
        \includegraphics[width=.48\linewidth]{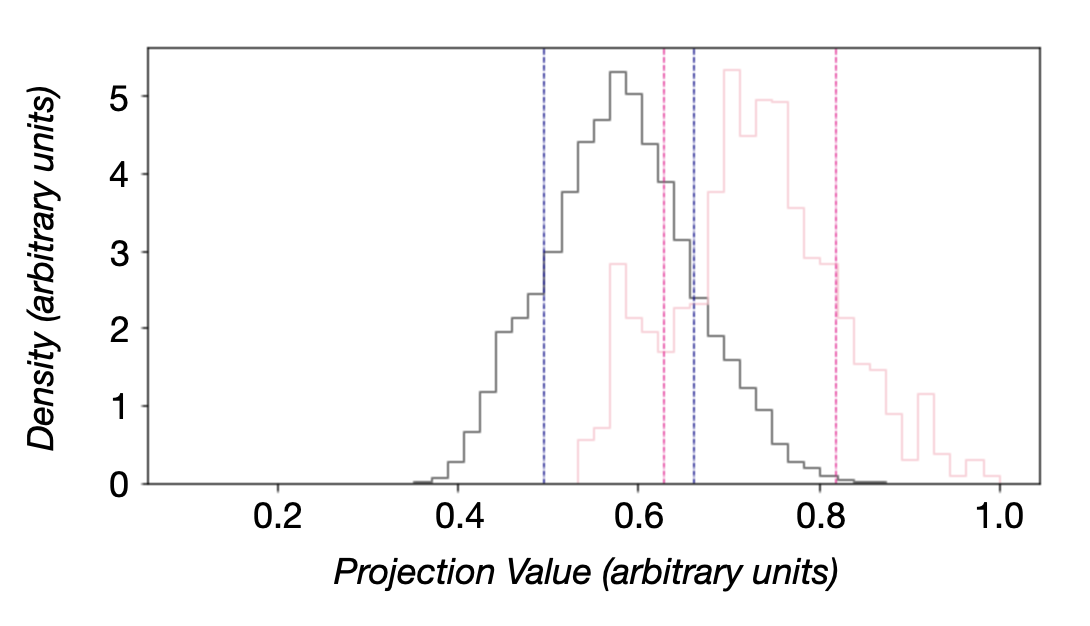}%
        \label{subfig:5c}%
    }\hfill
    \subfloat{%
        \includegraphics[width=.48\linewidth]{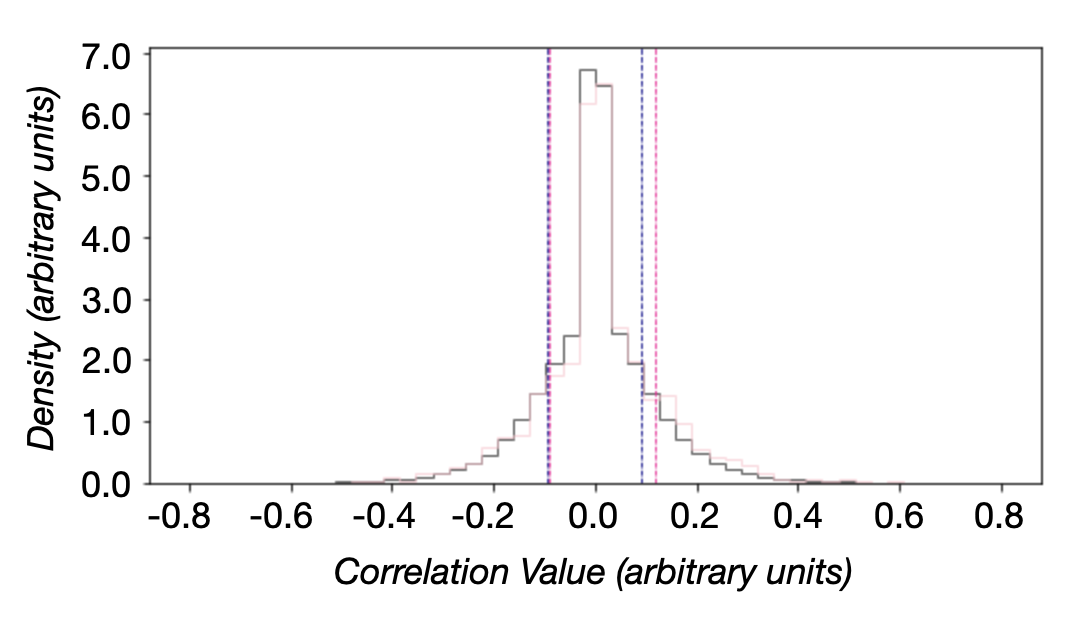}%
        \label{subfig:5d}%
    }
    \caption{A comparison of Doppler tomography (left) versus cross-correlation (right) for the recovery of the night-side carbon monoxide signal at 2.3 \micron, including cross-sections through $v_x$ = 0 km s$^{-1}$ and $v_{\mathrm{sys}}$ in the Doppler tomogram and CCF map, respectively, as well as the peak planetary velocity. Here the black dashed lines indicate the position of the peak velocity (see Table \ref{tab:Kps}) and the red dashed lines indicate the expected planetary velocity given by \protect\cite{Brogi2014}, ($K_p = 142.8$ km s$^{-1}$). The bottom panels show the difference in distribution of the `in-trail' (pink) and `out-of-trail' (black) signals for each method. Vertical lines indicate the 16-84 percentiles.}

    \label{fig:CO_Night}
    \end{figure*}

The successful detections of carbon monoxide and water within the atmosphere of HD 179949 b can be seen in Figures \ref{fig:Brogi_CO_ind} - \ref{fig:CO_Night}. We present the detection using Doppler tomography on the left, and the associated cross-correlated detection on the right. Cross-hairs indicate the expected $K_p$ \citep[red: taken from][]{Brogi2014} and the $K_p$ calculated in this work in black. Each plot has associated cross-sections through the peak signal, as well as through $v_{\mathrm{sys}}$ of the planetary system and $ v_x = 0$ km s$^{-1}$ for the cross-correlation projections and Doppler tomograms, respectively. These are the only cross-sections that are equivalent and are able to be directly compared between the two methods (see Section~\ref{sec:DT}). Though the cross-sections through $v_y$ and $K_p$ at the planetary orbital velocity in the Doppler tomograms and the CCF projections should not be directly compared due to the highly differing methodologies, we have provided them to give a further idea of the background noise levels seen in both methods. The intensities of these signals are normalised to assist with this comparison. In addition we also show the histograms produced from the `in-trail' (pink) and `out-of-trail' (black) distributions using both methods. Here the 16-84\% values are indicated by the vertical lines.

Where a planetary signal was detected in the CCF and/or Doppler maps, we fit these with a 2D Gaussian in order to determine $K_p$ values for the recoveries. The associated errors on these measurements are given by the 1$\sigma$ uncertainty on the 2D Gaussian fit. The $K_{p}$ values of each of the signals presented in this work can be seen in Table \ref{tab:Kps}.

We calculate the significance of the molecular features using several methods. We first calculate the signal-to-noise directly from both the CCF and and Doppler maps. In addition, we are able use the Welch's t-test on the cross-correlated values to determine a significance between the signal in- and out- of the planetary trail. While this, in general, cannot be calculated using the planetary trails generated from the Doppler maps, here we have elected to calculate the significance from the difference in 16-84\% ranges for both the in-trail and out-of-trail distributions (as described in Section \ref{sec:uncertainies}). We are also able to calculate this for the cross-correlation values. In order to perform a comparison between the two methods we have presented each of these values in Tables \ref{tab:results1} and \ref{tab:results_DT}.

Finally, in the case of the Doppler maps we measured any shift in the signal from the $v_{x} = 0$ km s$^{-1}$ line, which indicates a potential orbital phase shift from the predicted value. This enables us to update the orbital ephemeris for HD 179949 b (see Appendix). The Doppler tomograms presented in this paper have been corrected for the orbital phase shift and hence the planetary signals appear on the $v_{x} = 0$ km s$^{-1}$ line. The original measured offsets in $v_x$ are listed in Table~\ref{tab:results1} for completeness.

\begin{table*}
\caption{The planetary orbital velocity, $K_p$, of each detection within this work calculated by applying a 2D Gaussian fit to the CCF and Doppler tomography (DT) maps. The final column lists the $v_x$ values measured from the Doppler maps before updating the orbital period (see Section~\ref{sec:phasing}). All the non-zero $v_x$ values listed here are consistent with the expected orbital phase slip prior to this orbital period correction.}
\label{tab:results1} 
\begin{tabular}{cccccccc}
\hline
Molecule  & Wavelength  &  $K_p$ (CCF: km s$^{-1}$) & $v\mathrm{_{sys}}$ (CCF: km s$^{-1}$) & $K_p$ (DT: km s$^{-1}$) & $v_x$ (DT: km s$^{-1}$)  \\ 

\hline CO (Dayside)  & 2.3 \micron   & 146.7 $\pm$ 9.0  & -23.7 $\pm$ 1.1 &  141.8 $\pm$ 3.2 & 1.2 $\pm$ 1.2 \\

 H$_2$O   & 2.3 \micron  & 139.6 $\pm$ 9.8 & -25.7 $\pm$ 0.8 & 137.7 $\pm$ 7.7 & -1.6 $\pm$ 0.5 \\
 
 H$_2$O   & 3.5 \micron   & 138.1  $\pm$ 7.9 & -26.0 $\pm$ 1.8 & 134.2 $\pm$ 5.1 & 1.0 $\pm$ 0.9 \\
 H$_2$O   & 2.1 \micron   & 141.9 $\pm$ 3.4 & -25.1 $\pm$ 3.4 &  146.6 $\pm$ 3.6  & -6.3 $\pm$ 0.9 \\
 CO (Nightside) & 2.3 \micron  & 142.6 $\pm$ 9.5 & -26.5 $\pm$ 4.2 & 142.5 $\pm$ 6.2 & -0.3 $\pm$ 5.0 \\
\label{tab:Kps}
\end{tabular}
\end{table*}

\begin{table*}
\caption{The statistical significance for each detection found via CCF analysis. Several methods have been applied to attempt to provide a comparison with that found via Doppler tomography.}\label{tab:results_CCF} 

\begin{tabular}{cccccccc}
\hline
Molecule  & Wavelength ($\mu$m) & CCF SNR & CCF Projection & CCF Welch's T-Test  \\ 
\hline 
CO (Dayside)  & 2.3   & 5.5 & 0.047 & 6.3 \\
H$_2$O   & 2.3   & 5.0  & 0.047  & 6.4 \\
H$_2$O   & 3.5   & 7.9  & 0.159  &  8.5  \\
H$_2$O   & 2.1   & 5.7  & 0.070  & 5.4  \\
CO (Nightside) & 2.3   & 4.9  & 0.068 & 4.5  \\

\end{tabular}
\end{table*}

\begin{table*}
\caption{The statistical significance for each detection found via Doppler tomography. Several methods have been applied to attempt to provide a comparison with that found via cross-correlation. }\label{tab:results_DT} 
\begin{tabular}{cccccccc}
\hline
Molecule & Wavelength ($\mu$m) & DT SNR & DT Projection \\ 
\hline 
CO (Dayside)  & 2.3  & 91.5  & 7.9  \\
H$_2$O   & 2.3 & 185.9 & 6.1  \\
H$_2$O   & 3.5   & 87.8 & 3.5  \\
H$_2$O   & 2.1   & 138.5 &  9.1\\
CO (Nightside) & 2.3  & 19.6  & 1.2 \\

\end{tabular}
\end{table*}

\subsection{Day-side carbon monoxide and water at 2.3 \micron}
\label{sec:Dayside_23}

Day-side CO and H$_2$O detections in HD 179949 b have previously been reported by both \cite{Brogi2014} and \cite{Webb2020}. In particular, \cite{Brogi2012} reported a detection of CO (with SNR = 5.8) and a weaker detection of water (SNR = 3.9) at 2.3\micron\,with a joint SNR of 6.3 and $K_p$ of 142.8 $\pm$ 3.4 km s$^{-1}$. While \cite{Brogi2014} elected to use the 1$\sigma$ drop in the Welch's t-test to determine the errors on the planetary velocity, we cannot apply this method to our Doppler tomography analysis. As a result, each of the $K_p$ errors quoted within the text is calculated using a 2D Gaussian fit in order to remain internally consistent.

Our Doppler and CCF maps of the 2.3\micron\, region assuming either a CO- or a H$_2$O-only model/linelist are presented in Figures~\ref{fig:Brogi_CO_ind} \&~\ref{fig:Brogi_H2O_ind}, respectively. In all cases we recover a signal at the expected location of the planet. For the CCFs we recover the CO and H$_2$O with a SNR of 5.5 and 5.0, respectively. The SNR for CO is consistent with that found by \cite{Brogi2014}, while our water detection has a higher significance. We attribute this to our use of SYSREM to remove systematics and telluric contamination, while \cite{Brogi2014} applied simpler airmass detrending methods. By fitting a 2D Gaussian to the CCFs, we determine $K_p$ values of  146.7 $\pm$ 9.0 km s$^{-1}$ and  139.6 $\pm$ 9.8 km s$^{-1}$ for CO and H$_2$O, respectively. These are consistent with previous detections performed by \cite{Brogi2014}.

The equivalent Doppler maps are shown in the left panels of Figures~\ref{fig:Brogi_CO_ind}  and ~\ref{fig:Brogi_H2O_ind}. Using the same methodology as applied to the CCF maps, we determine a SNR of 91.5 for CO and 185.9 for water. However, as stated previously, such SNR estimates in the Doppler maps are misleading since the regularisation statistic results in highly non-Gaussian intensities in the maps --- which are generally more dominated by localised noise structures. Comparing the in-trail and out-of-trail distributions gives a significance value of 7.9 and 6.1 for CO and $\mathrm{H_{2}O}$, respectively.

Fitting a 2D Gaussian to the Doppler maps gives a $K_p$ of 141.8.7 $\pm$ 3.2 km s$^{-1}$ and  137.7 $\pm$ 7.7 km s$^{-1}$. Once again, these values are consistent with \cite{Brogi2014}, as well as with the CCFs calculated within this work. 

In both the CCF and Doppler maps we observe an offset between the $K_p$'s measured for CO and H$_2$O, with the water signal exhibiting an apparent lower orbital velocity. We note that the CCFs reported by \cite{Brogi2014} (see their Fig. 2) also exhibit a similar offset in $K_p$, though this is never explicitly discussed in that paper. We also see other offsets from the expected $K_p$ for H$_2$O in  other wavelength ranges -- we discuss these in Section~\ref{sec:conclusions}.






\subsection{Day-side water detection at 2.1\micron\,and 3.5\micron}
\label{sec:H2O_21_35}

We also searched for signatures of water at 2.1\micron\, and 3.5\micron. The detection of water at 3.5\micron\,for HD 179949 b was published by \cite{Webb2020}, who reported a weak (SNR = 4.8) signal. We also recover H$_2$O at 3.5\micron\,in both our CCF and Doppler maps (see Figure~\ref{fig:H2O_35}), but at higher signal-to-noise (SNR = 7.9 \& 87.4, respectively - though note the previous caveat regarding SNR estimates for Doppler tomography). Again, the higher SNR for the CCF detection reported in this work is likely due to our use of SYSREM to remove systematic effects, which was not applied by \cite{Webb2020}. We determine a significance of 3.5 for the Doppler tomography detection by comparing the in-trail and out-of-trail distributions. We find $K_p$ values of 138.1 $\pm$ 7.9 km s$^{-1}$ and 134.2 $\pm$ 5.1 km s$^{-1}$ from CCF and Doppler tomography analysis, respectively.

On the other hand, there has been no successful detection of water reported in the literature at 2.1\micron\,for HD 179949 b. As can be seen in Figure~\ref{fig:H2O_21}, we recover a signal in both the CCF and Doppler maps with SNRs of 5.7 and 138.5, respectively, and a significance of 5.4 and 9.1 for the CCF and Doppler maps, respectively. To the best of our knowledge, this is the first detection of water at this wavelength for this planet (though it was previously searched for by \citealt{Barnes2008}). This is likely due to advances in both telluric correction and data analysis techniques.

The CCF gives a $K_p$ value of 141.9 $\pm$ 3.4 km s$^{-1}$, while the Doppler map gives a $K_p$ of 146.6 $\pm$ 3.6 km s$^{-1}$. Unlike the 2.3 \& 3.5\micron\,water detections, which both show a $K_p$ offset to values lower than the expected planetary orbital motion, the 2.1\micron\, data does not show such an offset. We discuss this further in Section~\ref{sec:discussion}. 


\subsection{2.3 \micron\, - night-side carbon monoxide}
\label{sec:Nightside}

As noted in Section~\ref{sec:intro_planet}, simulations by \cite{deKok2014} showed that it should be feasible to detect emission from the night-side of HD 179949 b, though this has not been observationally achieved to-date for any exoplanet. In order to test this, we have also applied our analysis to the night-side spectra and searched for the presence of CO. The results are shown in Figure~\ref{fig:CO_Night}, which show a weak detection of CO in the night-side emission spectrum in the CCF analysis, with a more clear signal in the Doppler tomogram. Both detected signals appear at the expected velocity of the planet with $K_p$ = 142.6  $\pm$  9.5 $\mathrm{kms^{-1}}$ and 142.5  $\pm$  6.2  $\mathrm{kms^{-1}}$ for the CCF and Doppler maps, respectively. We believe this is the first ever observational evidence for molecular absorption in the night-side emission spectrum of an exoplanet. For the CCF analysis, the strongest signal (with a SNR = 4.9) was recovered for a model atmosphere with $T_2$ = 800 K, $T_1$ = 1000 K, $P_1$ = 1$\times 10^{-2}$ bar, and a VMR = 1$\times 10^{-6}$. We also attempted to detect  H$_2$O on the night-side, but were unsuccessful. This does not necessarily mean that water is lacking in this region, but due to the low significance of our detection of CO, it may well be the case that water falls below our detection threshold. Therefore, while the absence of a water detection in this region could be caused by a natural process, such as poor re-circulation, we do not wish to speculate on the physical implications of this.

\subsection{HCN, CH$_4$, and CO$_2$ non-detections}
\label{sec:Non_detections}

We investigated the presence of planetary signatures from HCN, CH$_4$, and CO$_2$. No signs of these molecules were apparent in either the CCF or Doppler maps for the range of atmospheric model parameters detailed in Table~\ref{tab:pt}.

\subsection{Discussion}
\label{sec:discussion}

For each of the detections presented, Doppler tomography gives clearer maps and sharper planetary signals compared to CCFs. This, however, comes at the cost of less straight-forward statistics and therefore requires a more complex approach to assessing the robustness of detected signals. We have outlined a process whereby we back-project the signal from the Doppler map to create a planetary trail, comparable to that produced in the CCF analysis. This allows us to generate a significance from the in-trail and out-of-trail distributions. While we have not been able to provide a single direct one-to-one comparison between the two detections it is clear from the combination of the images of the maps, the signal-to-noise calculations and the stark differences within the `in-trail' and `out-of-trail' distributions that Doppler tomography effectively enhances Doppler spectroscopy signals.

We note that the uncertainties on the measured $K_p$'s in the Doppler tomograms are, in general, lower than those determined for the CCFs when fitting a 2D Gaussian. One possible reason for this is the strict positivity criterion imposed by the regularisation statistic in Doppler tomography. This means that if the continuum is not correctly fit then only the sharper cores of the planetary lines may be revealed. However, we believe that this is not the case given that this would lead to a weaker detection. Additionally, \cite{Watson2019} showed that Doppler tomography produced better constrained planet signals in carefully controlled simulations. These sharper peaks likely also help increase the significance of planetary detections since the planet light is smeared over less pixels in velocity-space. 

In the case of HD 179949 b, we find tentative evidence for an offset in the detected water signal to a lower $K_p$ than expected at 2.3 \& 3.5\micron, while there is little evidence for such an offset at 2.1\micron. This offset is also apparent, though not discussed, for the 2.3\micron\,water discovery by \cite{Brogi2014}, as well as possibly in the 3.5\micron\,detection reported by \cite{Webb2020}. \cite{Webb2020} tested different models that do appear to affect the recovered $K_p$ and structure of the H$_2$O detection in their CCFs. We therefore do not believe that these offsets are physical. Rather, they are likely an interplay between the exact model used, telluric contamination, and phase coverage.

\section{Conclusions}
\label{sec:conclusions}
We have presented an analysis of the atmosphere of HD179949 b using a novel technique: Doppler tomography. We have shown that Doppler tomography can reproduce detections of CO and H$_2$O at 2.3 and 3.5 \micron, providing an independent confirmation of these detections as well as significantly reducing the noise levels within the outputs. We have also presented a new detection of water at 2.1 \micron, as well as a tentative signal from CO originating from the night-side of the planet, representing the first detection of molecular absorption from the night-side emission from an exoplanet.

From this, we have shown that performing night-side observations of exoplanets with high-resolution Doppler spectroscopy is a viable option. The current implementation of Doppler tomography used here assumes no line strength variation along the planetary orbit, however future observations of planets throughout their entire phase curves may allow us to investigate variation in abundance and temperature throughout its orbit. First steps towards this type of analysis have already been made using cross-correlation (e.g. \citealt{Herman2022}, \citealt{VanSluijs2022}). Implementing this within Doppler tomography will further boost our ability to study the dynamics of planetary systems. 

We also note that current Doppler tomography analysis relies on simple line lists, with only the line positions and scales, instead of the full radiative transfer models used in cross-correlation analysis. Further work will be done to incorporate these atmospheric models into the Doppler tomography code, which will only act to further improve retrievals using Doppler tomography. 

While we have implemented the use of an injected atmospheric signal, in order to determine the `optimal' number of SYSREM iterations to detect the planetary signal, we note that there is evidence that this will bias towards a detection. \cite{Cabot2020} found that in ~30\% of cases this resulted in a spurious signal. It is then important to highlight that, while we have presented the best case scenario for each of these signals, it is only through repeat detection that they can be verified. The tentative night-side detection within HD 179949 b would be an ideal candidate for follow-up using CRIRES+.

\newpage

\section*{Acknowledgements}
C.A.W. and E.dM. would like to acknowledge support from the UK Science and Technology Facilities Council (STFC, grant number ST/X00094X/1).
M.B., D.S. and T.R.M. also acknowledge support from the UK STFC via grant ST/T000406/1.\newline

\section*{Data Availability}
The data used within this article is presented in Table \ref{tab:obs}. All data is available from the ESO archive with the given ESO ID.

\bibliographystyle{mnras}
\bibliography{Biblo} 

\appendix
\section{Appendix}
\label{sec:appendix}
\subsection{SYSREM optimisation with RMS}
While each of the CCFs and Doppler maps that are presented above were generated using the optimal number of SYSREM iterations calculated from injection tests, we recognise that this can bias results in favour of a detection.

\begin{figure}
\centering

	\includegraphics[width = 8cm]{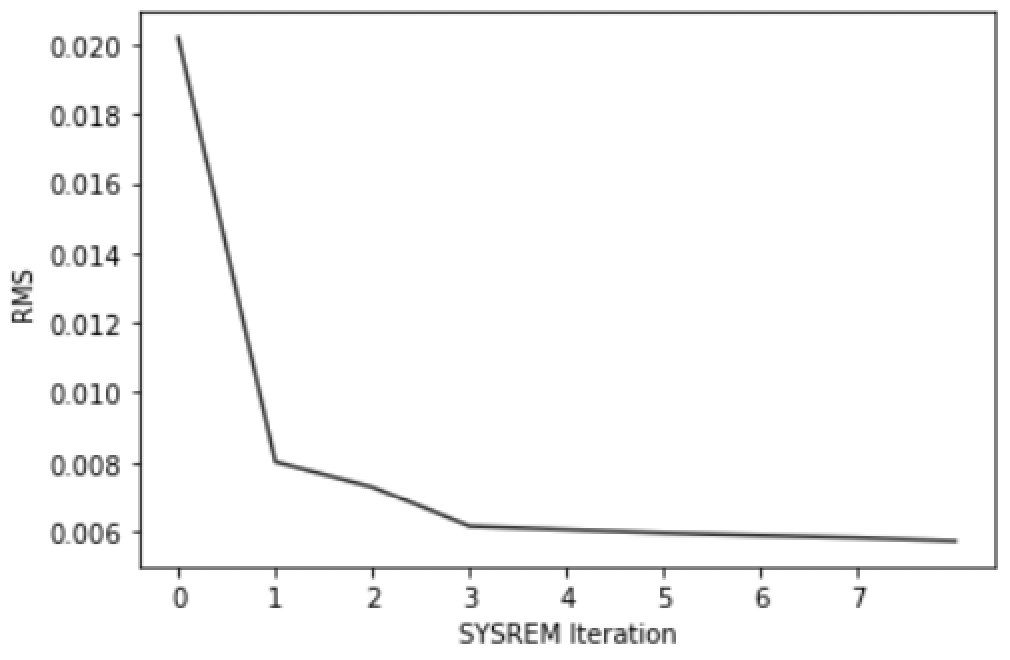}
    \caption{The RMS after each SYSREM iteration, calculated for detector 2 on July 19 2011. From this we can determine that three iterations is optimal, as after this the change in RMS does not decrease significantly with each additional iteration.}
    \label{fig:RMS}
\end{figure}

In order to validate these detections we have also calculated the optimal number of SYSREM iterations using the RMS of the data. This method, as applied in \cite{Herman2020}, avoids bias towards a given model, while still optimising noise removal within the data. An example of this calculation can be seen in Figure \ref{fig:RMS}. Here we see that the majority of change in the RMS occurs within the first three iterations, after which the value levels off and continues to decrease only slightly. From this we can determine that the optimal number of iterations for this particular data-set is three.

This RMS calculation was used to determine the optimal number of iterations for each detector used in the data presented above. While this method does lead to increased noise levels within the resulting Doppler maps and CCFs (typical CCF SNR ratio drops $\sim1\sigma$) in each case the signal persists, adding a level of validity to our above results.

\begin{figure*}
    \subfloat{%
        \includegraphics[width=.48\linewidth]{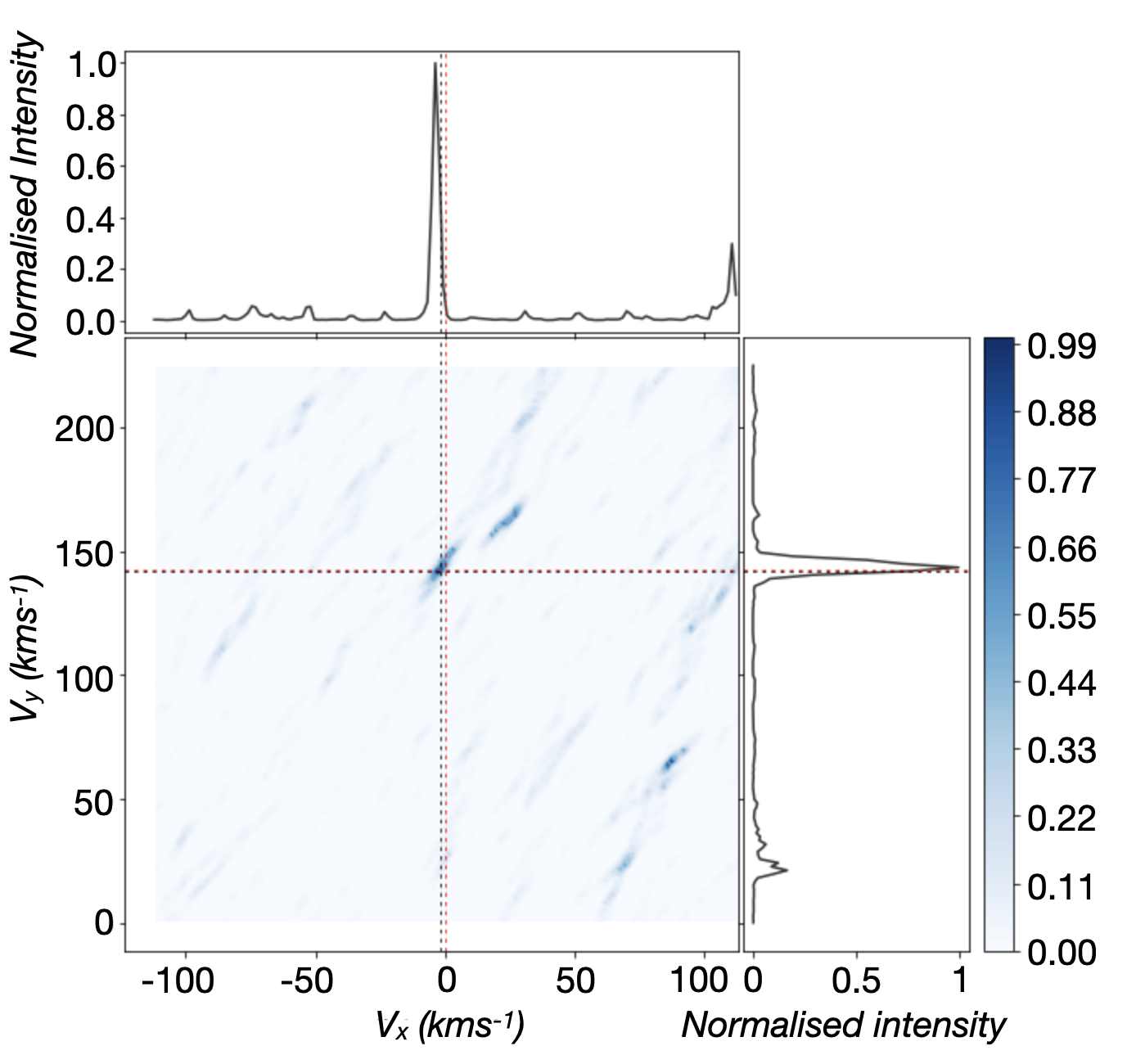}%
        \label{subfig:6a}%
    }\hfill
    \subfloat{%
        \includegraphics[width=.48\linewidth]{HD17_CO_23_Day_DT_Updated_CAW.png}%
        \label{subfig:6b}%
    }\\

    \caption{A comparison of the Doppler tomograms produced for the dayside 2.3\micron\ carbon monoxide detections using both the RMS method (left) and the injection test method (right). While the general background noise levels remain low in both, the RMS detection suffers from larger noise spikes. However, in both cases the planetary signal remains clear.}

    \label{fig:CO_comp}
    \end{figure*}

In general, by using this method we found the optimal number of SYSREM iterations lay in the range of 2-4, with the highest number applied being 6.

\begin{table}
\caption{The phase shift that is applied to each set of observations in order to centre the planetary signal on the expected $v_x$ = 0 km s$^{-1}$ line.}
\label{tab:phase_correction}
\begin{tabular}{ccc}
\hline
Observation Date (MJD) & $\Delta v_x$ (km s$^{-1}$)  & $\Delta \phi$ \\ 
\hline 
 54311.260367 & -6.3 $\pm$ 0.9 & 0.00700 $\pm$ 0.00100\\
 55768.263894 & -1.6 $\pm$ 0.5 & 0.00180 $\pm$ 0.00056 \\
 56813.984065 & 1.0 $\pm$ 0.9 & -0.00100 $\pm$ 0.00090\\
 
\end{tabular}
\end{table}

\subsection{Updated orbital ephemeris}
\label{sec:phasing}
Due to the non-transiting nature of HD 179949 b, the time of conjunction ($T_0$ ) and orbital period ($P$) of the planet are not as well-constrained or as frequently measured as that of a transiting exoplanet. Often we have to rely on measurements that are several years old when attempting to perform high-resolution Doppler spectroscopy on non-transiting planets. For example, \cite{Brogi2012}, who observed HD 179949 b in 2011, had to rely on a $T_0$ measurement last taken in 2006 \citep{Butler2006}. While the initial error on these values is indeed small, over the course of several years, they can compound to significant levels, resulting in a phase shift from the expected value. Since the planetary phase angle is vital for the CCF calculations, a small deviation in the phase will serve to weaken the peak signal, while large deviations will render it undetectable. This is something that has been previously observed, with phase shifts applied to observations of HD 179949 b \citep{Brogi2012} in order to maximise the CCF detections.

Within a Doppler map, a similar phase shift does not weaken the planetary signal. While we would normally expect the planetary signal at some point along the $v_x$ = 0 km s$^{-1}$ line, in the case of a planetary phase shift, it will instead appear shifted at an angle away from this line, in proportion to the shift in planetary phase. This is extremely useful as it gives a direct measurement of this phase shift at a given time for a set of observations, which can be used to calculate an updated ephemeris and orbital period.

Using the Doppler maps presented previously, we determined the required orbital phase shifts needed to centre the detected planetary signal onto the $v_x$ = 0 km s $^{-1}$ line. The measured phase offsets were calculated by fitting a 2D Gaussian to the original Doppler tomogram. These offsets are presented in Table~\ref{tab:phase_correction}, from which we calculated an updated ephemeris. The phase shifts that are associated with these offsets are within the 1$\sigma$ uncertainty on the orbital phase of HD 179949 b ($\Delta\phi$ = 0.012, \citet{Butler2006}).

This updated ephemeris was used for the analysis presented in the main body of this work.

\subsection{Planetary trails}
In order to produce a statistical comparison between the CCF and Doppler tomography techniques, we have generated a back-projected planetary trail from the final Doppler maps. This allows us to compare the distributions of pixels that are within the expected planetary trail (i.e. the signal), and those that are not (i.e. noise). To generate a significance from these trails we compare the 16-84\% ranges for both the ‘in-trail’
and ‘out-of-trail’ distribution.

Here we present the planetary trails from both Doppler tomography and CCF analysis. Both have been shifted into the planetary rest frame, appearing along the $v_{x}$/$v_{\mathrm{sys}}$ = 0 km s$^{-1}$ line.

We note that there are secondary streaks visible alongside the planetary trails in the H$_{2}$O Doppler tomgraphy results within Figures \ref{fig:H2O_23_Trail} and \ref{fig:H2O_21_Trail}. While we are unsure as to the origin of these features, be they a real planetary feature or a systematic, one possible cause could be changes in the line profile over the course of the orbit. This is an effect that has been predicted from General Circulation Models \citep[e.g.][]{Beltz2022}.

\begin{figure*}
    \subfloat{%
        \includegraphics[width=.48\linewidth]{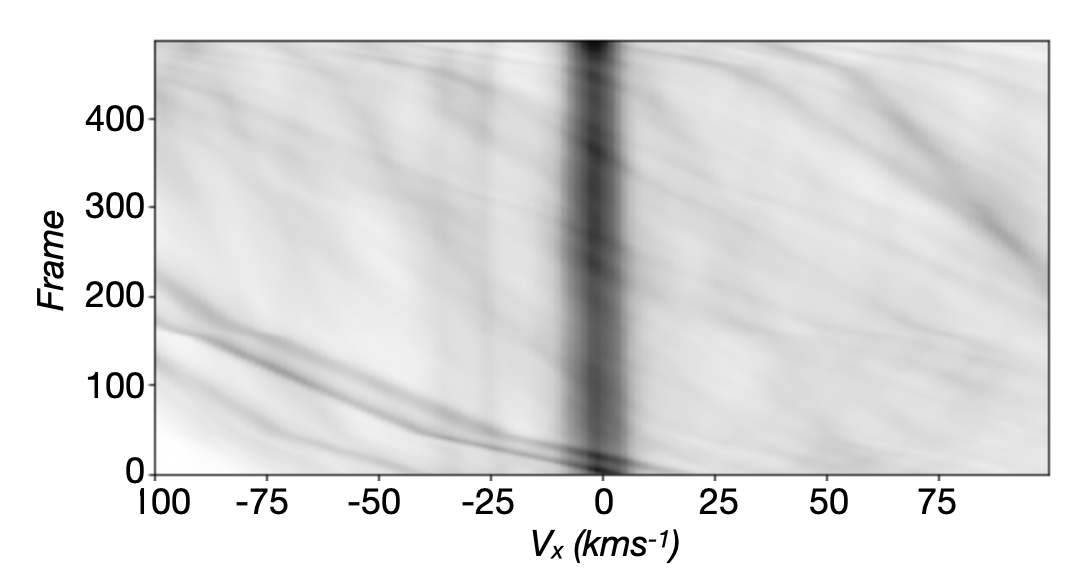}%
        \label{subfig:7a}%
    }\hfill
    \subfloat{%
        \includegraphics[width=.48\linewidth]{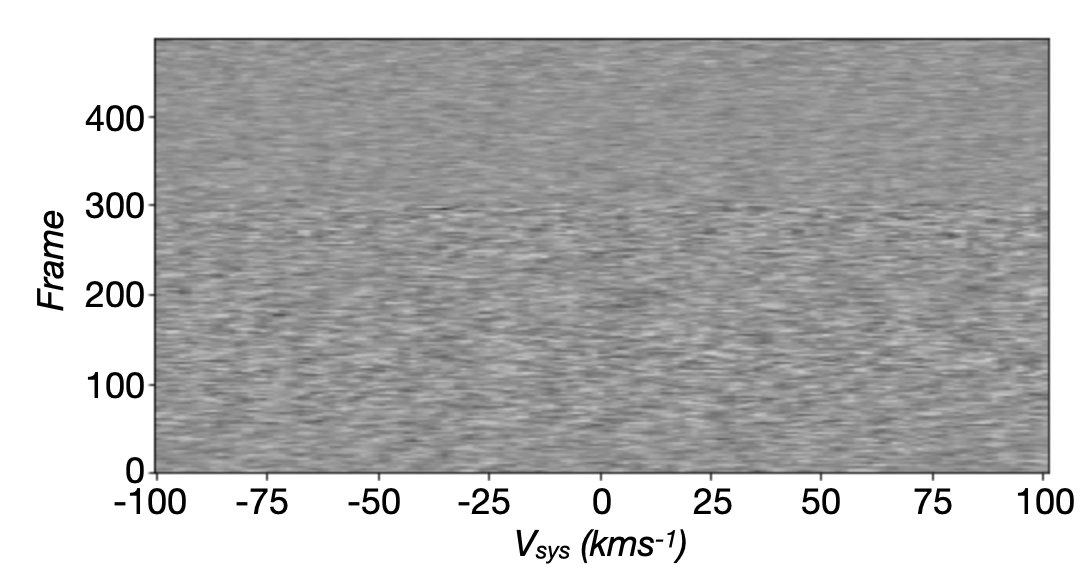}%
        \label{subfig:7b}%
    }\\
    
    \caption{A comparison of the projected trail from the Doppler tomography analysis of the dayside CO signal at 2.3\micron\ (left) vs. the cross-correlation values for the same signal. The expected planetary trail is along the $v_x$/$v_\mathrm{sys}$   = 0 km s$^{-1}$ line for each. }
    \label{fig:CO_23_Trail}
    \end{figure*}

\begin{figure*}
    \subfloat{%
        \includegraphics[width=.48\linewidth]{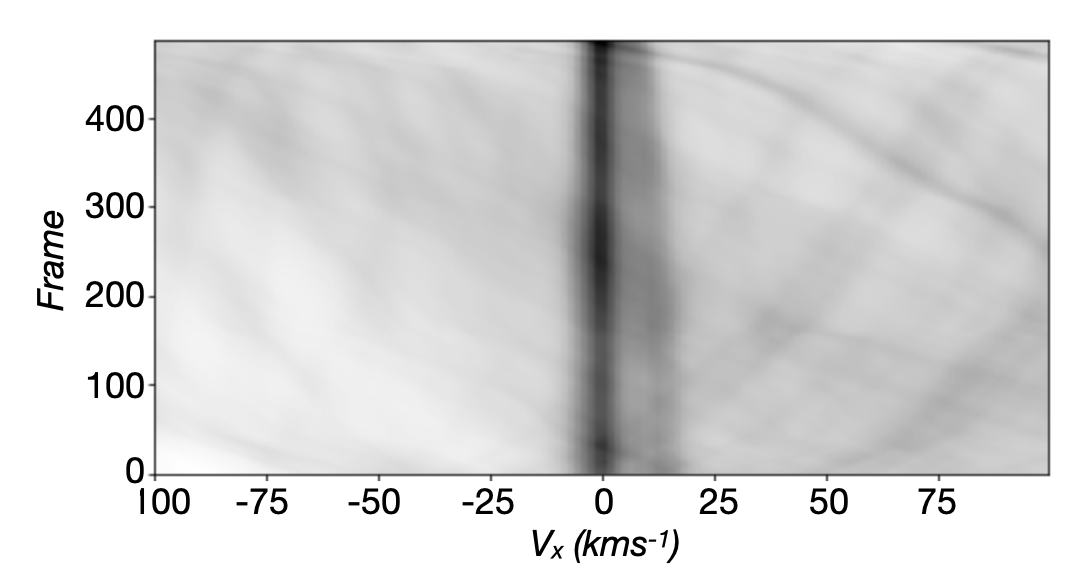}%
        \label{subfig:8a}%
    }\hfill
    \subfloat{%
        \includegraphics[width=.48\linewidth]{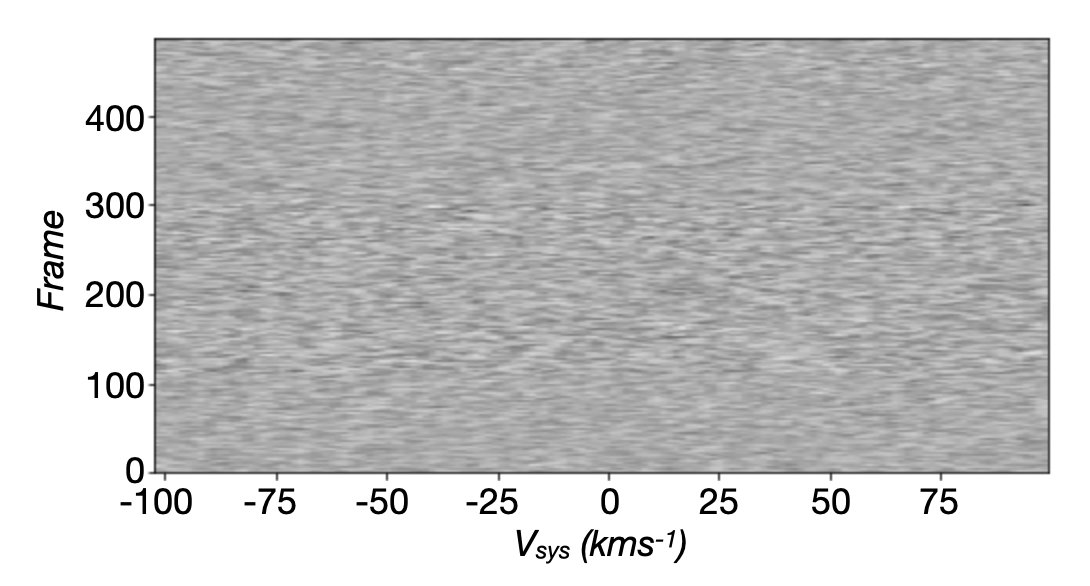}%
        \label{subfig:8b}%
    }\\
    
    \caption{A comparison of the projected trail from the Doppler tomography analysis of the dayside H2O signal at 2.3\micron\ (left) vs. the cross-correlation values for the same signal. The expected planetary trail is along the $v_x$/$v_\mathrm{sys}$   = 0 km s$^{-1}$ line for each. }

    \label{fig:H2O_23_Trail}
    \end{figure*}

\begin{figure*}
    \subfloat{%
        \includegraphics[width=.48\linewidth]{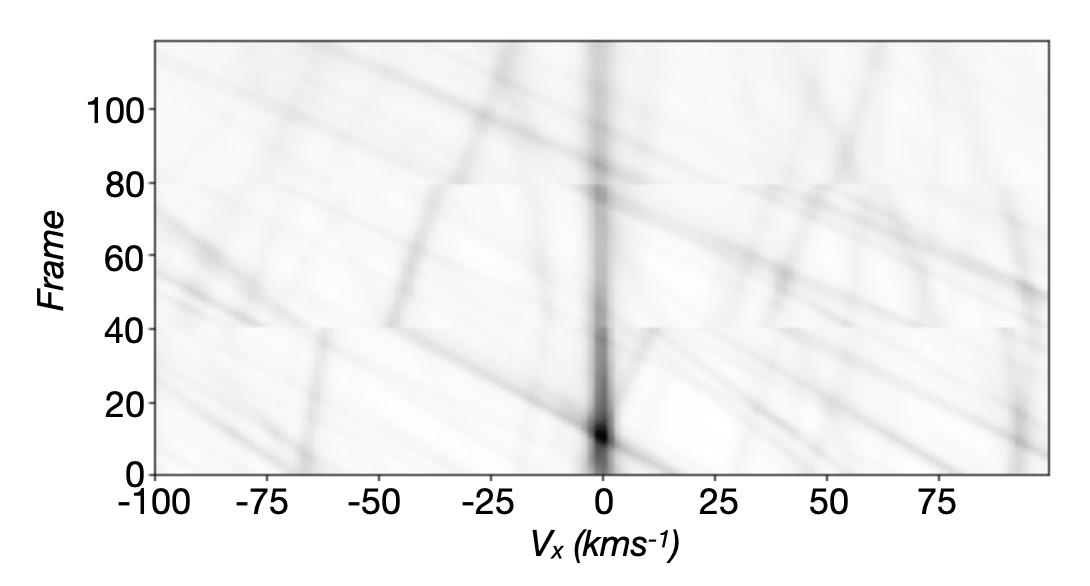}%
        \label{subfig:9a}%
    }\hfill
    \subfloat{%
        \includegraphics[width=.48\linewidth]{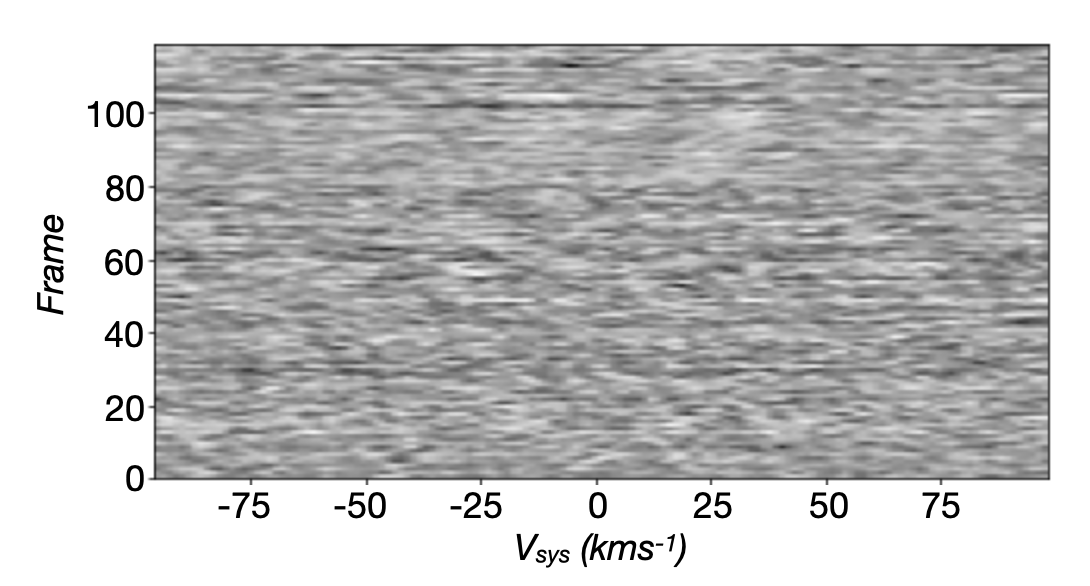}%
        \label{subfig:9b}%
    }\\
    
    \caption{A comparison of the projected trail from the Doppler tomography analysis of the dayside H2O signal at 3.5\micron\ (left) vs. the cross-correlation values for the same signal. The expected planetary trail is along the $v_x$/$v_\mathrm{sys}$  = 0 km s$^{-1}$ line for each. }

    \label{fig:H2O_35_Trail}
    \end{figure*}

\begin{figure*}
    \subfloat{%
        \includegraphics[width=.48\linewidth]{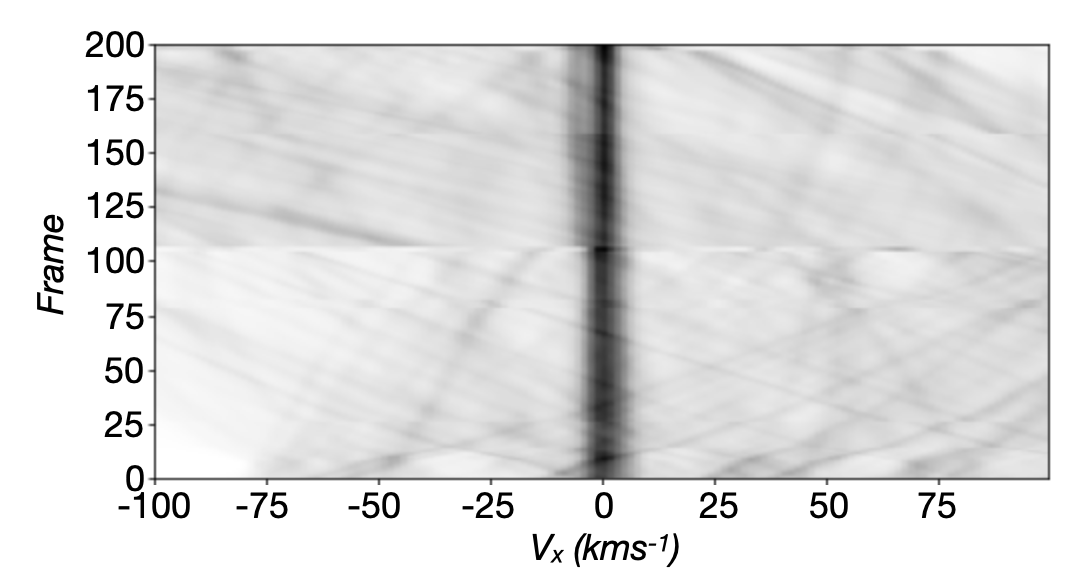}%
        \label{subfig:10a}%
    }\hfill
    \subfloat{%
        \includegraphics[width=.48\linewidth]{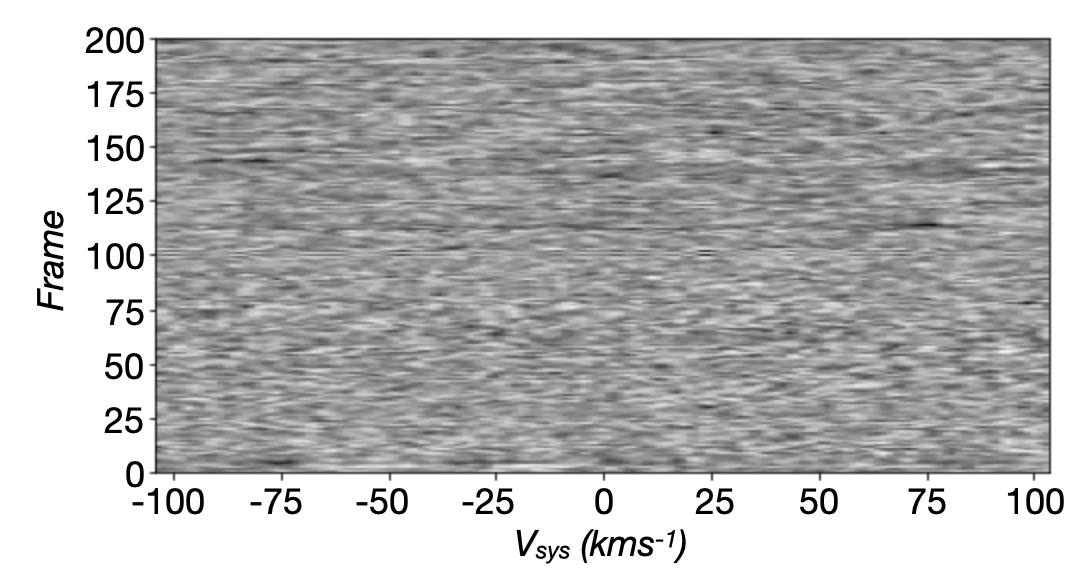}%
        \label{subfig:10b}%
    }\\
    
    \caption{A comparison of the projected trail from the Doppler tomography analysis of the dayside H2O signal at 2.1\micron\ (left) vs. the cross-correlation values for the same signal. The expected planetary trail is along the $v_x$/$v_\mathrm{sys}$ = 0 km s$^{-1}$ line for each.}

    \label{fig:H2O_21_Trail}
    \end{figure*}

\begin{figure*}
    \subfloat{%
        \includegraphics[width=.48\linewidth]{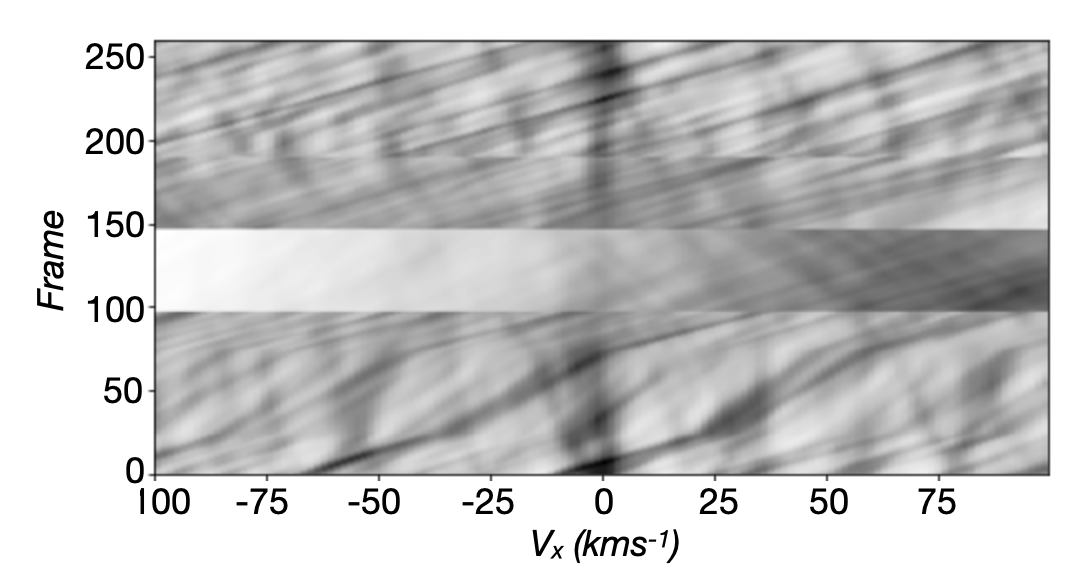}%
        \label{subfig:11a}%
    }\hfill
    \subfloat{%
        \includegraphics[width=.48\linewidth]{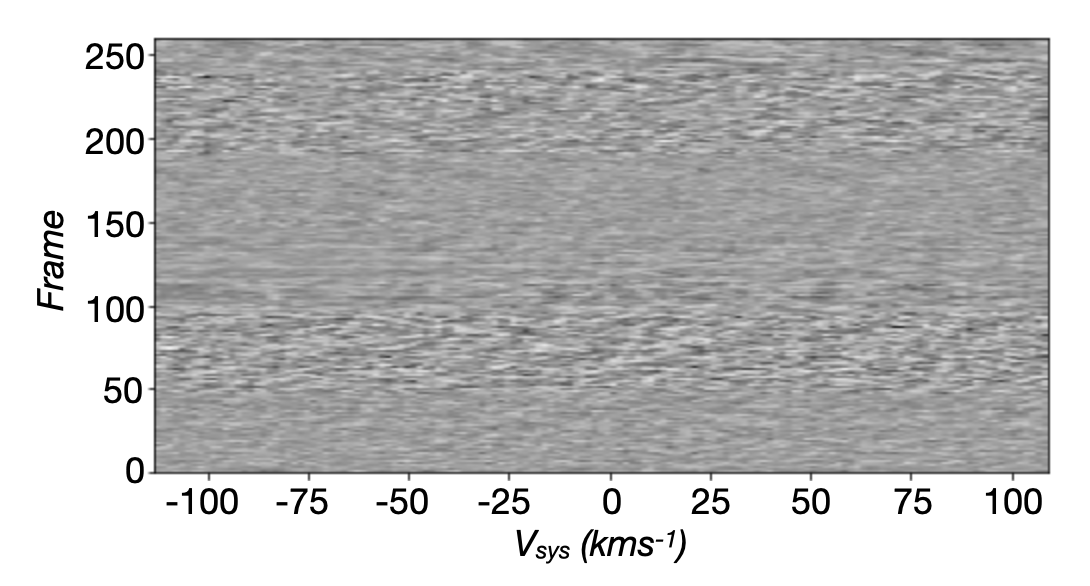}%
        \label{subfig:11b}%
    }\\
    
    \caption{A comparison of the projected trail from the Doppler tomography analysis of the nightside CO signal at 2.3\micron\ (left) vs. the cross-correlation values for the same signal. The expected planetary trail is along the $v_x$/$v_\mathrm{sys}$   = 0 km s$^{-1}$ line for each.}

    \label{fig:CO_23_Night_Trail}
    \end{figure*}

\bsp	
\label{lastpage}
\end{document}